\documentclass[preprints,article,accept,pdftex,oneauthors]{Definitions/mdpi} 
\firstpage{1} 
\pubvolume{1}
\issuenum{1}
\articlenumber{0}
\pubyear{2026}
\copyrightyear{2026}
\datereceived{ } 
\daterevised{ } 
\dateaccepted{ } 
\datepublished{ } 
\hreflink{} 
\usepackage[english]{babel}

\usepackage{bm}        

\usepackage{physics}

\theoremstyle{plain}

\usepackage{mathtools}
\usepackage[normalem]{ulem}
\usepackage{soul}

\renewcommand{\expval}[1]{{\left\langle#1\right\rangle}}
\renewcommand{\dd}{{\rm d}}
\newcommand{\imi}{{\rm i}}

\renewcommand{\dv}[2]{\frac{\dd #1}{\dd #2}}
\newcommand{\dhv}[2]{{\dd #1}\big/{\dd #2}}

\newcommand{\Boltz}{ k_{\rm\scriptscriptstyle B}}

\renewcommand{\dim}{{\rm dim}}
\renewcommand{\Tr}{{\rm Tr}}

\newcommand{\J}{\mathbb{J}}
\newcommand{\smallbbJ}{{\scalebox{0.6}{$\mathbb{J}$}}}
\newcommand{\M}{\mathbb{M}}
\newcommand{\Jbar}{{\overline{\mathbb{J}}}}
\newcommand{\ddt}[1]{{\frac{\displaystyle {\rm d}#1}{\displaystyle {\rm d}t}}}
\newcommand{\inlineddt}[1]{{{\rm d}#1/{\rm d}t}}

\newcommand{\A}{{\mathbf A}}

\newcommand{\half}{{\textstyle\frac{1}{2}}}
\newcommand{\Hil}{{\mathcal H}}
\newcommand{\PHES}{{\mathcal P}_{\rm HES}(\Hil)}

\renewcommand{\J}{{\mathbb{J}}}
\renewcommand{\A}{{\mathbb{A}}}
\newcommand{\B}{{\mathbb{B}}}

\newcommand{\smallK}{{\scalebox{0.6}{$K$}}}
\newcommand{\smallL}{{\scalebox{0.6}{$L$}}}
\newcommand{\smallM}{{\scalebox{0.7}{$M$}}}
\newcommand{\smallN}{{\scalebox{0.7}{$N$}}}

\newcommand{\smallequalone}{{\scalebox{0.7}{$=1$}}}
\newcommand{\PH}[1]{{P_{\scriptstyle #1}}}

\newcommand{\DrhoJ}{{\mathcal{D}^\J_\rho}}
\newcommand{\HiliK}{{\Hil_{\scriptstyle i_\smallK}^{\scriptstyle \smallK}}}
\newcommand{\PK}{{P_{\scriptstyle \smallK}}}
\newcommand{\PJ}{{P_{\scriptstyle \smallL}}}
\newcommand{\HK}{{H_{\scriptstyle \smallK}}}

\newcommand{\HKJ}{H_{\scriptstyle \smallK_\J}^{\smallbbJ}}

\newcommand{\SK}{S_{\scriptstyle \smallK}}
\newcommand{\MK}{{M_{\scriptstyle \smallK}}}
\newcommand{\SKJ}{{S_{\scriptstyle \smallK_\J}^{\smallbbJ}}}
\newcommand{\SKJHE}{{S_{\scriptstyle \smallK_\J}^{\smallbbJ,\text{HE}}}}
\newcommand{\SKJtildeHE}{{\tilde S_{\scriptstyle \smallK_\J}^{\smallbbJ,\text{HE}}}}
\newcommand{\SKtilde}{{{\tilde S}_{\scriptstyle \smallK}}}
\newcommand{\SLtilde}{{{\tilde S}_{\scriptstyle \smallL}}}
\newcommand{\rhoKtilde}{{\tilde\rho_{\scriptstyle \smallK}}}
\newcommand{\rhoKJtildeHE}{{\tilde\rho_{\scriptstyle \smallK_\J}}^{\smallbbJ,\text{HE}}}
\newcommand{\rhoKJtilde}{{\tilde\rho_{\scriptstyle \smallK_\J}}^{\smallbbJ}}
\newcommand{\sK}{{s_{\scriptstyle \smallK}}}
\newcommand{\sKJ}{{s_{\scriptstyle \smallK_\J}^{\smallbbJ}}}
\newcommand{\ZK}{{Z_{\scriptstyle \smallK}}}
\newcommand{\ZL}{{Z_{\scriptstyle \smallL}}}
\newcommand{\ZKJ}{{Z_{\scriptstyle \smallK_\J}^{\smallbbJ}}}
\newcommand{\betaK}{{\beta_{\scriptstyle \smallK}}}
\newcommand{\betaL}{{\beta_{\scriptstyle \smallL}}}
\newcommand{\betaKJ}{{\beta_{\scriptstyle \smallK_\J}^\smallbbJ}}
\newcommand{\alphaKJ}{{\alpha_{\scriptstyle \smallK_\J}^\smallbbJ}}
\newcommand{\alphaK}{{\alpha_{\scriptstyle \smallK}}}
\newcommand{\tauK}{{\tau_{\scriptstyle \smallK}}}

\newcommand{\iJ}{{\scriptstyle i_\J}}
\newcommand{\PiK}{{P_{\scriptstyle i_\smallK}^{\scriptstyle \smallK}}}
\newcommand{\PiJ}{{P_{\scriptstyle i_\J}^{\smallbbJ}}}
\newcommand{\RiJ}{{R_{\scriptstyle i_\J}^{\smallbbJ}}}
\newcommand{\tauiJ}{{\tau_{\scriptstyle i_\J}^{\smallbbJ}}}
\newcommand{\Tau}{{\tilde\tau}}
\newcommand{\PiL}{{P_{\scriptstyle i_\smallL}^{\scriptstyle \smallL}}}
\newcommand{\PiKJ}{{P_{\scriptstyle i_{\smallK,\smallbbJ}}^{\scriptstyle \smallK,\smallbbJ}}}
\newcommand{\PKJ}{{P_{\scriptstyle \smallK_\J}^{\smallbbJ}}}
\newcommand{\pK}{{p_{\scriptstyle \smallK}}}
\newcommand{\pKSE}{{p_{\scriptstyle \smallK}^\text{SE}}}

\newcommand{\pKJ}{{p_{\scriptstyle \smallK_\J}^{\smallbbJ}}}

\newcommand{\piK}{{p_{\scriptstyle i_\smallK}^{\scriptstyle \smallK}}}

\newcommand{\piKJ}{{p_{\scriptstyle i_{\smallK,\smallbbJ}}^{\scriptstyle \smallK,\smallbbJ}}}
\newcommand{\piL}{{p_{\scriptstyle i_\smallL}^{\scriptstyle \smallL}}}
\newcommand{\giK}{{g_{\scriptstyle i_\smallK}^{\scriptstyle \smallK}}}
\newcommand{\xiK}{{x_{\scriptstyle i_\smallK}^{\scriptstyle \smallK}}}
\newcommand{\yiK}{{y_{\scriptstyle i_\smallK}^{\scriptstyle \smallK}}}

\newcommand{\eiJ}{{\varepsilon_{\scriptstyle i_\J}^{\smallbbJ}}}
\newcommand{\miJ}{{m_{\scriptstyle i_\J}^{\smallbbJ}}}
\newcommand{\miK}{{m_{\scriptstyle i_\smallK}^{\scriptstyle \smallK}}}
\newcommand{\eiK}{{\varepsilon_{\scriptstyle i_\smallK}^{\scriptstyle \smallK}}}

\newcommand{\ejK}{{\varepsilon_{\scriptstyle j_\smallK}^{\scriptstyle \smallK}}}

\newcommand{\eiKJ}{{\varepsilon_{\scriptstyle i_{\smallK,\smallbbJ}}^{\scriptstyle \smallK,\smallbbJ}}}
\newcommand{\siK}{{s_{\scriptstyle i_\smallK}^{\scriptstyle \smallK}}}
\newcommand{\wiK}{{w_{\scriptstyle i_\smallK}^{\scriptstyle \smallK}}}

\newcommand{\sjK}{{s_{\scriptstyle j_\smallK}^{\scriptstyle \smallK}}}
\newcommand{\iKJ}{{i_{\scriptstyle \smallK,\smallbbJ}}}
\newcommand{\tauKJ}{{\tau_{\scriptstyle \smallK_\J}^{\scriptstyle \smallbbJ}}}

\newcommand{\tauiKJ}{{\tau_{\scriptstyle i_{\smallK,\smallbbJ}}^{\scriptstyle \smallK,\smallbbJ}}}
\newcommand{\sumN}{{\sum\nolimits_{\scriptstyle\,i\smallequalone}^{\,\smallN}\,}}
\newcommand{\sumNJ}{{\sum\nolimits_{\scriptstyle\,i_\J\smallequalone}^{\,\smallN_\J}\,}}
\newcommand{\SumNJ}{{\sum_{\scriptstyle\,i_\J\smallequalone}^{\,\smallN_\J}\,}}
\newcommand{\sumK}{{\sum\nolimits_{\scriptstyle\,\smallK\smallequalone}^{\,\smallM}\,}}
\newcommand{\sumiK}{{\sum\nolimits_{\scriptstyle\,i_\smallK\smallequalone}^{\,\smallM_\smallK}\,}}
\newcommand{\sumiKJ}{{\sum\nolimits_{\scriptstyle\,i_{\smallK,\smallbbJ}\smallequalone}^{\,\smallM_{\smallK,\smallbbJ}}\,}}
\newcommand{\SumK}{{\sum_{\scriptstyle\,\smallK\smallequalone}^{\,\smallM}\,}}

\newcommand{\SumKJ}{{\sum_{\scriptstyle\,\smallK_\J\smallequalone}^{\,\smallM_\J}\,}}
\newcommand{\sumKJ}{{\sum\nolimits_{\scriptstyle\,\smallK_\J\smallequalone}^{\,\smallM_\J}\,}}
\newcommand{\SumiK}{{\sum_{\scriptstyle\,i_\smallK\smallequalone}^{\,\smallM_\smallK}\,}}
\newcommand{\SumiKJ}{{\sum_{\scriptstyle\,i_{\smallK,\smallbbJ}\smallequalone}^{\,\smallM_{\smallK,\smallbbJ}}\,}}
\newcommand{\sumL}{{\sum\nolimits_{\scriptstyle\,\smallL\smallequalone}^{\,\smallM}\,}}
\newcommand{\sumiL}{{\sum\nolimits_{\scriptstyle\,i_\smallL\smallequalone}^{\,\smallM_\smallL}\,}}
\newcommand{\x}{{\mathbf{x}}}

\definecolor{rkrPurple}{HTML}{73024F}

\definecolor{rkrblue}{HTML}{13b2ba}

\Title{Evolution of Hypoequilibrium States in Steepest Entropy Ascent Models for Nonequilibrium Quantum Thermodynamics}

\TitleCitation{Hypoequilibrium  and  Steepest Entropy Ascent}

\Author{Gian Paolo Beretta $^{1}$\orcidA{}*, Rohit Kishan Ray $^{2}$\orcidB{}, and Michael R. von Spakovsky $^{3}$ \orcidC{}}

\AuthorNames{Gian Paolo Beretta, Rohit Kishan Ray, and Michael R. von Spakovsky}

\isAPAStyle{%
       \AuthorCitation{Beretta, G.P., Ray, R.K., von Spakovsky, M.R.}
         }{%
        \isChicagoStyle{%
        \AuthorCitation{Beretta, Gian Paolo; Ray, Rohit Kishan; and von Spakovsky, Michael R.}
        }{
        \AuthorCitation{Beretta, G.P.; Ray, R.K.; von Spakovsky, M.R.}
        }
}

\address{%
\textsuperscript{1} \quad Department of Mechanical and Industrial Engineering, University of Brescia, 25123 Brescia, Italy; gianpaolo.beretta@unibs.it\\
\textsuperscript{2} \quad Department of Materials Science and Engineering, Virginia Tech, Blacksburg, VA 24061, USA; rkray@vt.edu\\
\textsuperscript{3} \quad Department of Mechanical Engineering, Virginia Tech, Blacksburg, VA 24061, USA; vonspako@vt.edu}

\corres{Correspondence: gianpaolo.beretta@unibs.it}

\abstract{A formal development of the hypoequilibrium (HE) state concept within the Steepest-Entropy-Ascent Quantum Thermodynamics (SEAQT) framework is presented, emphasizing its rigorous mathematical formulation. Using a general decomposition of the Hilbert space, HE states are defined in operator language and the reduced evolution of the associated intensive parameters for the regime where the dissipative dynamics commutes with the Hamiltonian is derived. It is proved that the $M$-th order HE family (where $M$ is the number of spectral sectors) constitutes an invariant manifold under the SEAQT equation of motion, ensuring that states initially representing a “mixture of canonicals” maintain this structure throughout their evolution. Furthermore, a formal connection is established between the HE ansatz and the rate-controlled constrained equilibrium (RCCE) method, identifying HE variables as constraint potentials. Finally, the model is extended to non-Hamiltonian SEAQT (NH-SEAQT) interactions to describe thermodynamically consistent energy and entropy exchanges between subsystems and heat baths. This work provides the formal foundation for reduced-order modeling of far-from-equilibrium relaxation and transport processes, and supports a methodology previously applied across various physical and chemical systems.}


\begin{document}

\section{Introduction}\label{sec:intro}

An equation of motion able to model nonequilibrium relaxations at any scale must balance two competing demands, namely: (i) be sufficiently microscopic to respect the laws of conservation as well as  the second law of thermodynamics, yet (ii) have a sufficiently reduced description in terms of a small number of physically interpretable variables. Without the latter, it becomes unusable for systems with large state spaces, while the former suggests that it should be thermodynamically self-consistent, i.e., able to evolve from any initial nonequilibrium non-zero-entropy state to stable equilibrium.

Within the unified quantum theory of mechanics and thermodynamics proposed by Hatsopoulos and Gyftopoulos~\cite{hatsopoulos_1976_unified, hatsopoulos_1976_unifieda, hatsopoulos_1976_unifiedb, hatsopoulos_1976_unifiedc}, the two  steepest-entropy-ascent (SEA) equations of motion proposed for nonequilibrium quantum thermodynamics (QT) in \cite{Beretta1981,Beretta1984,berettaQuantumthermodynamicsnew1985} and developed in several papers thereafter (see, e.g., \cite{Beretta2009,Beretta2014,Beretta2020,RayBeretta2025}) satisfy the first of these demands. These SEAQT evolution equations, one for an unstructured quantum system and the other for a structured composite of quantum subsystems, augment the unitary quantum evolution of the von Neumann equation of motion \cite{BreuerPetruccione_2007_TheTheory,Neumann_1955_MathematicalFoundations} with a dissipative term constructed so that entropy production is nonnegative, while the relevant dynamical invariants (e.g., normalization and energy and, when present, other commuting generators of the motion) are preserved by the dissipative part of the motion. In its compact quantum form, each equation evolves the density operator $\rho$ through a nonlinear law governed by a generalized Massieu-type operator, which encodes both the entropy and relevant conservation constraints~\cite{Beretta2009,Beretta2014,Beretta2019}. The resulting dynamics is thermodynamically structured and thermodynamically self-consistent, but, nonetheless, a dynamics on the full space of density operators~\cite{Korsch1987,Hensel1992,Gheorghiu2001a,Gheorghiu2001b,Tabakin2017,Tabakin2023}.
	

	Note that the standard equation of motion widely used by the quantum thermodynamics community is the Kossakowski-Lindblad equation \cite{Lindblad_1976_OnThe} or more precisely the Gorini-Kossakowski-Sudarshan-Lindblad (GKSL) equation  \cite{GoriniKossakowskiSudarshan_1976_CompletelyPositive, kosloff_2019_Quantumthermodynamics, manzano_2020_short}.  The basis of this equation is the generator of a completely positive dynamical semigroup, which results in a large class of quantum master equations for which Kossakowski provides the necessary and sufficient generator conditions \cite{Kossakowski_1972_OnNecessary} and Ingarden and Kossakowski the consistency condition for the macroscopic observables of open quantum systems  \cite{IngardenKossakowski_1975_OnThe}. An equivalent approach is the Kraus-operator or operator-sum formalism  \cite{Kraus1983,NielsenChuang_2009_QuantumComputation,Choi_1975_CompletelyPositive} of trace-preserving linear completely positive definite maps introduced by Kraus  \cite{Kraus_1971_GeneralState,Stinespring_1955_PositiveFunctions}. Unlike the SEA equations of motion, which are \textit{inherently} quantum mechanically and thermodynamically self-consistent, the GKSL equation is only \textit{inherently} the former since as pointed out by Spohn \cite{Spohn_1976_ApproachTo, spohn_2025_irreversible}, whether or not this equation evolves thermodynamically  depends on the choice of Kraus operators for which the corresponding semigroup has a unique final stable equilibrium (Gibbs) state to which every initial state evolves. Thus, the dissipative path predicted by the GKSL equation is a thermodynamic path only if it obeys the ``unital'' condition \cite{Witten_2020_AMini} and that depends on the Kraus operators employed. In contrast, the dissipation operator of the SEAQT equations, which is constructed based on the SEA principle and a set of operators called the generators of the motion, always evolves towards stable equilibrium independent of the choice of these operators. This is what is meant by being \textit{inherently} thermodynamically self-consistent. An example of this is seen in the application of both formalisms to the evolution of slightly perturbed Bell diagonal states given in Chapter 12 of \cite{SpakovskyReynoldsAscencio2026}.

Now, regarding the second demand, using the equations of motion for systems with large state spaces becomes impractical due to the high dimensionality of the microscopic description unless a low-dimensional manifold can be identified that is both (i) physically meaningful and (ii) dynamically consistent with the SEA evolution. The hypoequilibrium (HE) concept provides such a principled reduction~\cite{Li2016a,Li2016b,Li2016c,Li2017,Li2018a,Li2018b}. Using this concept, the Hilbert space is decomposed into an orthogonal direct sum of a finite number of coarse spectral sectors $\{\Hil_{\smallK}\}_{\smallK=1}^M$. Each sector is the subspace spanned by a subset of the eigenspaces of the Hamiltonian operator, for which the total probability and mean energy (together with other moments if needed) are tracked. In this representation, each nonequilibrium state is described as a mixture of the density operators associated with the coarse spectral sectors, weighted by the respective total probabilities. When the Hilbert-space decomposition is generated by partitioning the set of energy eigenvalues into contiguous energy intervals, the coarse spectral sectors reduce to what are commonly called  ``energy shells.'' The HE approximation  then postulates that the density operator for each spectral sector is canonical (or grand canonical), and, thus, has its own inverse temperature (or temperature and chemical potential) different from the other spectral sectors. Each state of the system is, therefore, represented by a set of  ``spectral-sector temperatures'' (or ``spectral-sector temperatures and chemical potentials'') and the corresponding set of   ``spectral-sector total probabilities.'' This is not merely a numerical convenience. It is instead a way of encoding far-from-equilibrium structure while retaining an equilibrium-like form locally on each coarse spectral sector. The  HE concept for quantum systems  parallels the lower-dimensional-manifold-approximation philosophy  of several model reduction approaches, notably, the ``rate-controlled constrained-equilibrium'' (RCCE) method in  chemical kinetics \cite{Keck1971,Beretta2012,Hadi2015}, the ``quasi-equilibrium'' approximation in  dynamical systems \cite{Gorban2006,Karlin2016} and in complex biological models \cite{Snowden2017}  (characterized by a variety of empirical \cite{Keck1979,Keck1990} and geometrical \cite{Espanol2004,Pope2006,Grmela2014,Pavelka2020,Barbaresco2022,Morrison2024,Barbaresco2025} methods to identify the most effective slow invariant manifold for each given class of problems), or systematic coarse-graining techniques \cite{Chiavazzo2012,Beretta2016,Rivadossi2018,Voth2022}. 

Of course, to be physically relevant, the HE manifold of constrained-equilibrium (or quasi-equilibrium) states must be  invariant  with respect to the SEA dynamics. In other words, when evolved according to the SEA dynamics, any initial HES must remain within the HE manifold at all times. This dynamical consistency is important since otherwise the evolution risks being an uncontrolled approximation. If indeed it is consistent (under clearly stated assumptions) as suggested by the proof provided in \cite{Li2016a}, then HE--SEAQT model becomes a legitimate reduced model with the SEA principle providing the irreversible dynamics and the HE concept providing the reduced-order description.

 The HE concept has also been applied to the SEAQT equation for a general structured composite of quantum subsystems  \cite{Holladay2019} and to ad hoc extensions of the SEAQT equation tailored to model heat,  diffusion, and heat-and-diffusion interactions between systems as well as provide non-Hamiltonian (NH) extensions of the Onsager reciprocity relations to the non-near-equilibrium nonlinear domain as well as models of system--bath interactions \cite{Li2016b,Li2016c,Li2017}. Applications include (see \cite{SpakovskyReynoldsAscencio2026}) predicting electrical, thermal, magnetic, and mechanical transport properties; the chemical and electrochemical kinetics of reacting mixtures; ferromagnetic eddy current losses; nonequilibrium size and concentration effects on heat and mass diffusion; the kinetics of surface adsorptions and contamination; discontinuous and continuous phase decompositions; ordering and phase separation; atomistic spin relaxations; thermal expansions; microstructural evolutions; cell membrane lipid diffusion; defect formation and migration; the nonequilibrium behavior of nonquasiequilibrium thermodynamic cycles, etc.

In this paper,  the focus is on providing a precise and detailed mathematical formulation of the HE--SEAQT model for the simplest unstructured quantum systems and of the NH--HE--SEAQT model of non-Hamiltonian heat interactions between unstructured systems. For these models, the HE subspace is proven to be an invariant-manifold within the SEA dynamics.
Excluded from the present treatment is a discussion of variations of the NH--HE--SEAQT model such as those used in \cite{Holladay2019,Morishita2023a,Morishita2023b,Rochasoto2025,Damian2026,Rochasoto2026} to heuristically model the time evolution of coherences and correlations as well as energy, entropy, and particle exchanges between subsystems in the presence of interaction Hamiltonians. These variations involve hybrid equations of motion including the simultaneous effects of the von Neumann, Lindblad, and SEAQT terms as well as a continuous projection of the state operator onto a HE manifold.

In the present work, hypoequilibrium states (HES) are first formulated in compact operator language consistent with the SEAQT equation of motion for a simple quantum system, using a general decomposition of the Hilbert space and the corresponding HES variables. Next, based on the simplifying but important practical regime in which the dissipative SEA term acts on states commuting with the Hamiltonian (so that the dynamics reduces to a closed evolution of eigenlevel populations), the reduced evolution for the HE parameters is derived and the associated entropy production structure identified. Addressing the invariant-manifold property next, an initial $M$-th order HES representing a ``mixture of canonicals'' is used to show how the SEA evolution preserves the order of the initial state, demonstrating that the evolving states remain within the same $M$-th order HE family. The HE representation is also shown to lie within the logic of the RCCE~\cite{Keck1971,Keck1979,Keck1990} model reduction concept, since the HES can be viewed as a maximum-entropy state relative to a chosen set of coarse constraints so that the evolving intensive parameters acquire the meaning of constraint potentials. Without shifting the focus away from the HES--SEA dynamical consistency, the RCCE connection unifies terminology and situates the HE concept within a broader literature on reduced-order modeling.

The paper is organized as follows.  Section~\ref{sec:energy_rearrangement} defines the tenets of the HE concept and describes the Hilbert space decomposition induced by a specific partitioning of the set of energy eigenvalues. Section \ref{sec:hes} formally defines the HE ansatz, derives the reduced evolution in terms of HES variables, and isolates the minimal relations needed for later proofs. The HE formalism is then extended to structured composite systems in section~\ref{sec:CSHE}. Section \ref{sec:rcce} follows with a discussion of the consistency of the HE concept with the RCCE approach. Section~\ref{sec:heseaqt_single} outlines the implementation of the HE approximation into the original SEAQT dynamical structure in terms of the notation reviewed for completeness and consistency in Appendices \ref{sec:seaqt} and \ref{sec:heseaqt} and proves that the HE subspace is an invariant-manifold within the SEA dynamics. Section~\ref{sec:heseaqt_heat} then introduces a modification of the original SEAQT equation of motion for composite systems that models, in a thermodynamically consistent manner, energy and entropy (and mass) exchanges between subsystems as non-Hamiltonian (NH-SEAQT) dissipative effects. This is followed by Section~\ref{sec:NH2}, which shows the compatibility of the NH-SEAQT model with the notion of a heat interaction between two systems, while  Section~\ref{sec:NH3} formulates the model for a system in contact with two other systems that could model heat baths. Section \ref{sec:conclusions} then presents final conclusions.

\section{Preliminary HE Assumptions: Partitioning the Energy Spectrum}\label{sec:energy_rearrangement}

The technical assumptions underlying the HE concept are stated first. Only  density operators $\rho$ on the system Hilbert space $\Hil$ that belong to  the special class defined by the following conditions are considered: 
\begin{enumerate}[label=\textbf{(HE\arabic*):},     labelsep=4pt, leftmargin=*]
    \item $\rho$ commutes with $H$ (at all times $t$), i.e., $[H,\rho]=0$; 
    \item $\rho$  is full rank, i.e., has no zero eigenvalues, so that $P_{\rho>0}=I$, where $P_{\rho>0}$ is the projection operator onto the range of $\rho$; 
    \item $\rho$ assigns equal probability to each of the $g_i$ corresponding eigenstates of each degenerate eigenvalue $\varepsilon_{i}$ of $H$. 
\end{enumerate}
To be more explicit, the $N$-energy-eigenlevel system Hamiltonian in spectral form is written as 
\begin{equation}
	H=\sumN \varepsilon_i \PH{i}\, ,
\end{equation} 
where 
$\varepsilon_i$ denotes the $i$-th distinct system energy eigenvalue with degeneracy $g_i=\dim\Hil_{i}=\Tr(\PH{i})$ ($\dim\Hil=\sum_{i=1}^N g_i$); and $\PH{i}$ is the projector onto the corresponding eigenspace $\Hil_{i}$ so that $\Hil=\bigoplus_{i=1}^N\Hil_{i}$ is the spectral decomposition of  $\Hil$ induced by $H$ and $I= \sum_{i=1}^N\PH{i}$ is the corresponding resolution of the identity operator. With these assumptions, the density operator has the spectral form 
\begin{equation}
	\rho=\sumN p_i \PH{i}\, ,
\end{equation}
and $p_i$ is the occupation probability of the $i$-th energy eigenlevel and the entropy operator $S(\rho)$ defined according to~\cite{Beretta1981,Beretta1984,Beretta2009} is given by
\begin{equation}
	\label{eq:SandexpectS}
	S=-\Boltz P_{\rho>0}\ln\rho=-\Boltz\sumN (\ln p_i) \PH{i}= \sumN s_i \PH{i}\, ,
\end{equation}
where $\Boltz$ is the Boltzmann constant and 
 \begin{equation}
 	s_i=-\Boltz\ln p_i\, .
 \end{equation}
In general, for any trace preserving dynamics, i.e., with $\Tr(\inlineddt{\rho}) =0$,   $\langle\inlineddt{S(\rho)}\rangle = \Tr\left(\rho\, \inlineddt{S(\rho)}\right)=0$ (see footnote 7 of~\cite{Beretta2009} for a proof). 

The next assumption is: 
\begin{enumerate}[label=\textbf{(HE\arabic*):},start=4, labelsep=4pt, leftmargin=*]
\item The set of $N$  eigenvalues of $\rho$ is arbitrarily partitioned into $M$ disjoint subsets. 
\end{enumerate}
Each subset defines a subspace of the Hilbert space obtained as the span of the corresponding energy eigenspaces. This induces an orthogonal direct--sum decomposition of the Hilbert space  $\Hil$ so that the resolution of the identity $I$ and the spectral forms of $H$ and $\rho$ can be written as follows:
\begin{equation}\label{eq:I}
    \Hil=\bigoplus\nolimits_{\smallK =1}^\smallM \Hil_{\smallK} \, ,\qquad \Hil_{\smallK}=\bigoplus\nolimits_{i_\smallK =1}^{\smallM_\smallK} \HiliK\, ,
\qquad
    I=\sumK \PK  \, ,\qquad  \PK =\sumiK\PiK\, ,
\end{equation}
\begin{equation}\label{eq:H}
    H=\sumK \HK  \, ,\qquad  \HK =\sumiK \eiK  \PiK\, ,
\qquad
    \rho=\sumL \sumiL \piL  \PiL \, ,
\end{equation}
where clearly $ M_{\smallK} \geq  1$, $ \sumK M_{\smallK} = N $ and
\begin{equation}\label{eq:ortho_sub_proj}
	\PiL  \PiK =\delta_{\smallL\smallK}\delta_{i_\smallL i_\smallK}\PiK \,  ,\qquad \dim\HiliK=\Tr( \PiK)  =\giK \, ,
\end{equation}
\begin{equation}
\dim\Hil_{\smallK}=\Tr (  \PK) =g_\smallK= \sumiK\giK\, , 
\qquad 
\dim\Hil=\Tr(I)=\sumK    g_\smallK \, .\end{equation}

Following~\cite{Li2016a,Li2016b} and in anticipation of the additional assumptions introduced in Section~\ref{sec:hes}, the subspace $\Hil_{\smallK}$ is called the $K$-th hypoequilibrium spectral sector (HESS); and  the total probability $\pK$ and the mean energy $\langle \HK \rangle  $ of the $K$-th HESS are defined as
\begin{eqnarray}\label{eq:pK}
    \pK \!\!& =&\!\! \langle \PK \rangle= \Tr\left(\rho \PK\right) = \sumiK \langle \PiK \rangle = \sumiK  \piK \giK \, ,
    \\
    \label{eq:piK}
   p_{\scriptstyle\eiK}= \langle \PiK \rangle \!\!& =&\!\!  \Tr\left(\rho \PiK \right)= \sumL\sumiL \piL  \Tr(\PiL \PiK)=  \piK \giK \, ,
    \\ 
    \label{eq:EK}
    \langle \HK \rangle  \!\!& =&\!\! \Tr\left(\rho \HK \right) 
    =   \sumiK\piK \giK   \eiK \, ,
    \\ 
    \label{eq:EPK}
    \langle \HK  \HK \rangle \!\!& =&\!\! \Tr\left(\rho H^2_\smallK\right)
    =   \sumiK\piK \giK   (\eiK)^2\, ,
\end{eqnarray}
Here $\sum \pK =1$, and  the probability associated with the $i_\smallK$-th energy eigenlevel of the $\smallK $-th subset is not $\piK $ but $ p_{\scriptstyle\eiK}=\piK \giK$.

The $\smallK$-th \textit{HESS density operator} $\rhoKtilde $ on $\Hil_{\smallK}$ is defined by
\begin{equation}\label{eq:rho}
    \rho=\sumK\pK  \rhoKtilde\, ,  \qquad \rhoKtilde =\frac{1}{\pK } \sumiK \piK  \PiK \, ,
\end{equation}
where, by construction
\begin{equation}\label{eq:rhovHK}
		\rhoKtilde H_\smallL = H_\smallL\rhoKtilde =\delta_{\smallL\smallK} \HK\rhoKtilde \,  .
\end{equation}
To compute how each HESS contributes to the overall entropy, the logarithms of these density operators are needed. Defining
\begin{equation}\label{eq:siK}
    \siK =-\Boltz\ln \piK \, ,  \qquad \sK=-\Boltz\ln \pK  \, , 
\end{equation}
the \textit{proper} HESS entropy operator $\SKtilde $ and  the \textit{local} HESS entropy  $\langle \SKtilde\rangle$ are given by
\begin{equation}\label{eq:lnrhoK}
    \SKtilde=-\Boltz\ln\rhoKtilde=-\Boltz\sumiK\ln\frac{\piK }{\pK }\PiK = \sumiK(\siK -\sK)  \PiK\, , 
\end{equation}
\begin{equation}\label{eq:SK}
    \langle \SKtilde\rangle=\Tr\left(\rho \SKtilde\right) 
    =-\pK  \sK+\sumiK \piK \giK \siK\, ,
\end{equation}
\begin{equation} \label{eq:lnpikPik}
    \sumiK(\ln \piK )\PiK =(\ln \pK )\PK + \ln\rhoKtilde\, . 
\end{equation}
where Eq. (\ref{eq:rho}) and the fact that the projector $P_{i_K}^K$ is an idempotent operator (see \cite{Beretta2009}) are used to derive this last equation. Furthermore, the term \emph{local}  denotes a single HESS, \emph{i.e.}, a coarse spectral sector $\Hil_{\smallK}$; $\pK \sK$ is the \textit{partitional} HESS entropy; and  
\begin{equation}\label{eq:rhovSKtilde}
	\rhoKtilde \SLtilde = \SLtilde\rhoKtilde =\delta_{\smallL\smallK} \SKtilde\rhoKtilde \,  .
\end{equation}
Now, following the definition of the HESS and using Eq.~\eqref{eq:ortho_sub_proj}, 
\begin{equation}\label{eq:lnrho}
    \ln\rho=\sumK\sumiK(\ln \piK )\PiK 
    =\sumK [(\ln \pK )\PK + \ln\rhoKtilde ]\, ,
\end{equation}
\begin{eqnarray}\label{eq:Sop}
    S = -\Boltz\ln\rho=\sumK\sumiK\siK \PiK  =\sumK (\sK\PK +\SKtilde) =\sumK  S_\smallK\, ,
\end{eqnarray}
where the HESS \textit{partial} entropy operator, its expectation value, and its expected covariance with the \textit{local} Hamiltonian, $\HK$, are given by
\begin{equation}
    S_\smallK= \sK\PK +\SKtilde=\sumiK \siK\PiK\, ,
\end{equation}
\begin{equation}\label{eq:sk_avg}
	\langle \SK \rangle = \pK \sK  +\langle \SKtilde\rangle = \sumiK \piK \giK \siK\, , 
\end{equation}
\begin{equation}
	\langle \SK \HK\rangle = \Tr(\rho \HK \SK)  =  \sumiK \piK \giK \siK \eiK\, . 
\end{equation}
 The  HESS \textit{proper} energies and entropies are then expressed as
\begin{equation}\label{eq:EKK}
	\langle \HK \rangle_\smallK =\Tr\left(\rhoKtilde  \HK\right) 
	= \frac{1}{\pK }\sumiK\giK  \piK  \eiK =\frac{\langle \HK \rangle}{\pK }\, .
\end{equation}
\begin{eqnarray}\label{eq:SKK}
	\langle \SKtilde\rangle_\smallK&=&-\Boltz\Tr\left(\rhoKtilde \ln\rhoKtilde \right) = \Tr\left(\rhoKtilde  \SKtilde\right)  =-\Boltz\SumiK\left(\frac{\piK }{\pK } \ln\frac{\piK }{\pK }\right)\Tr(\PiK)\nonumber\\
	&=& \Boltz\ln \pK -\frac{\Boltz}{\pK }\SumiK\giK  \piK  \ln \piK=\frac{\langle \SKtilde\rangle}{\pK }= \frac{\langle  \SK \rangle}{\pK }-\sK\, .
\end{eqnarray}
A summary of the various energy and entropy definitions provided thus far is given in Table \ref{tab:energyentropy}.

\begin{table}
	\centering
	\caption{Summary of the various energy and entropy definitions associated with a coarse spectral sector  $\Hil_{\smallK}$ for a given partitioning of the energy spectrum, where $\langle X \rangle_\smallK=\Tr(\rhoKtilde X)$ and  $\langle X \rangle=\Tr(\rho X)=\sumK \pK \langle X \rangle_\smallK$.}
	\begin{tabular}{|l|c|c|c|c|}
		\hline 
		& \multicolumn{2}{c|}{\centering Energy} & \multicolumn{2}{c|}{\centering Entropy} \\ 
		\cline{2-5}
&  Operator &  Mean Value&  Operator &  Mean Value \\ 
		\hline 
		Proper & $\HK$ & $\langle \HK \rangle_\smallK=\Tr(\rhoKtilde  \HK) $ &  $\SKtilde$ & $\langle \SKtilde\rangle_\smallK= \Tr(\rhoKtilde  \SKtilde)$ \\ 
		\hline 
		Local & $\HK$ & $\langle \HK \rangle  =\pK \,\langle \HK \rangle_\smallK  $ &$\SKtilde$  & $\langle \SKtilde\rangle=\pK \,\langle \SKtilde\rangle_\smallK$  \\ 
		\hline 
		Partitional &$0$ & $0$ &  $\sK\PK$&  $\pK  \sK$ \\	\hline	
		Partial & $\HK$ & $\langle \HK \rangle$ & $\SK=\sK\PK + \SKtilde$ & $\langle  \SK \rangle=\pK  \sK+\langle \SKtilde\rangle_\smallK$ \\
		\hline 
		Overall & $\displaystyle H{=}\SumK  \HK $ & $\displaystyle \langle H\rangle{=}\SumK \langle \HK \rangle$ &  $\displaystyle S{=}\SumK  \SK$ & $\displaystyle \langle  S\rangle{=}\SumK \langle \SK \rangle$ \\ 
		\hline 
	\end{tabular} 
	\label{tab:energyentropy}
\end{table}

Although at first sight Assumption HE1 ($[H,\rho] =0$) may appear to be overly restrictive, it is a deliberate modeling choice with wide applicability for which HE is a statement about coarse-grained equilibration of eigenenergy populations. In particular, the defining signature of an HE sector is, as seen below (Eq.~\eqref{eq:siKHES}), that the sector entropic coordinates $\siK$ 
are affine maps on $\eiK$. Once the HES manifold is established and shown to be dynamically consistent in this minimal setting, one can ask how coherences and additional generators deform or enlarge that manifold. Furthermore, under the simplifying assumptions made in this section, a full description of the time evolution of the state operator $\rho$ still requires the time dependence of all the $\piK $'s or equivalently the $p_{\scriptstyle\eiK}$'s, whose number is $N$, and for practical systems may still be very large. Additional methods for dealing with this---such as the density of states method developed in \cite{Li2016a} and the approach used for phonons in \cite{Li2018b}---can be employed.

In the next two sections, a set of  assumptions is introduced to reduce the number of independent variables from $N$ to $2M$, where $M$ is the number of assumed HESS's. As shown in Section \ref{sec:rcce}, these assumptions are often physically justifiable and are conceptually aligned with a model-reduction strategy widely used in chemical kinetics.

\section{Main HE Assumption: Canonical HESS Density Operators}\label{sec:hes}

The HE approximation results when the HESS density operators are constrained to take a canonical or grand canonical form. The grand canonical case is not treated here. The formal assumption is as follows:
\begin{enumerate}[label=\textbf{(HE\arabic*):},start=5, labelsep=4pt, leftmargin=*]\item 
    A state $\rho$ is a ``HE state''  of order $M$ with respect to the chosen partition $\{\Hil_\smallK\}_{\smallK=1}^M$ if, for each HESS $K$, there exists an inverse temperature $\beta_\smallK$ such that
    \begin{equation}\label{eq:rhoKeq}
        \rhoKtilde^\text{HE} = \frac{\PK\exp(-\betaK  \HK  )\PK}{\ZK (\betaK )}, \qquad  \ZK (\betaK )=\Tr[\PK\exp(-\betaK  \HK )\PK]\, ,
    \end{equation}
    or, equivalently, using Eqs. (\ref{eq:H}) and (\ref{eq:rho})
    \begin{equation}\label{eq:piKeq}
        \frac{\piK }{\pK }= \frac{\exp(-\betaK  \eiK  )}{\ZK (\betaK )}, \qquad  \ZK (\betaK )=\SumiK \giK \exp(-\betaK  \eiK )\, .
    \end{equation}
\end{enumerate}
The kinetic justification of assumption HE5 is discussed in  Section~\ref{sec:rcce}. Note that Assumption \textbf{HE3} is a corollary of Assumption \textbf{HE5}. It is also noteworthy that the symmetric compression of the exponential terms via the projector $\PK$ in Eq.~\eqref{eq:rhoKeq} restricts the operator's support to the subspace  $\Hil_{\smallK}$. This ensures that $\rhoKtilde^\text{HE} $ acts nontrivially only within its respective sector. Without $\PK$, the term
would reduce to the identity on all other subspaces $\Hil_{\smallL\ne\smallK}$, leading to unphysical contributions in the global sum.

Combining Eqs.~\eqref{eq:rhoKeq} and~\eqref{eq:rho}, it is evident that the HE approximation consists of assuming that the states at any time are of the form
\begin{equation}\label{eq:rhohypo}
    \rho^\text{HE}=\SumK\pK 
    \frac{\PK\exp(-\betaK  \HK  )\PK}{\ZK (\betaK )}=\SumK\pK\,\rhoKtilde^\text{HE}\, .
\end{equation}
Note that, since by construction  the ranges of operators $\PK $ and $\HK$ are entirely contained in subspace $\Hil_{\smallK}$ and the  $\Hil_{\smallK}$'s are orthogonal to each other so that $\PJ\HK=\HK\PJ=\delta_{\smallL\smallK}\HK$, the HE density operator $\rho^{HE}$ can be compactly rewritten by absorbing the normalization factors and the sector probabilities into the exponential such that
\begin{equation}\label{eq:rhohypo2}
		\rho^\text{HE}=\SumK \PK
		\exp(-\alphaK \PK -\betaK  \HK  )\PK\, ,
\end{equation} 
where the coefficients $\alphaK$ are defined by the relation
\begin{equation}\label{eq:alphaK}
		\alphaK = \ln \ZK (\betaK )-\ln \pK =\ln \ZK (\betaK )+\sK/\Boltz\,.
\end{equation}
In this representation, $\alpha_K$ acts as a sector-dependent normalization constant (effectively a free energy shift) that ensures the correct weighting of each subspace.
The HE manifold is therefore characterized by the following equivalent relations
\begin{equation}\label{eq:pikHE}
	\piK=\exp(-\alphaK -\betaK  \eiK)  \,,
\end{equation} 
\begin{equation}\label{eq:sikHE}
	\siK=\Boltz\alphaK +\Boltz\betaK  \eiK  \, .
\end{equation} 
The set $\PHES$ of HE density operators of form Eq.~\eqref{eq:rhohypo} is an ``invariant manifold of the dynamics'' if the   underlying  equation of motion evolves any  density operator initially in $\PHES$ along a path that remains within $\PHES$  at all future times (``strongly'' invariant if this holds also backwards in time as is later proven to be the case under HE--SEAQT dynamics; see Eq.~\eqref{eq:inv_manifold}). A sufficient condition for this to hold true is that the equation of motion entails rates of change of the $\piK$'s compatible with Eq.~\eqref{eq:pikHE}, i.e., that the following
\begin{equation}\label{eq:pikdotHE}
	\dv{\piK}{t}=-\exp(-\alphaK -\betaK  \eiK)\,\left(\dv{\alphaK}{t} +\dv{\betaK}{t}  \eiK\right)  \,,
\end{equation}
or, equivalently,
\begin{equation}\label{eq:sikdotHE}
\frac{1}{\Boltz}	\dv{\siK}{t}=\dv{\alphaK}{t} +\dv{\betaK}{t}  \eiK  \, ,
\end{equation}
are satisfied.

By exploiting the fact that the $\PiK$ are mutually orthogonal projectors that resolve the identity $\PK$ within the subspace $\Hil_{\smallK}$, the exponential terms in Eq.~\eqref{eq:rhohypo2} can be decomposed into a sum of spectral components such that
\begin{align}
	\rho^\text{HE}&= {\displaystyle\SumK \PK\exp\Big[-\!\!\SumiK(\alphaK+\betaK\eiK)\,\PiK\Big] \PK} \\ &= 
{\displaystyle\SumK	\SumiK \exp\big(-\alphaK-\betaK\eiK\big)\,\PiK} \label{eq:exp_rho_hes} \\ &=
{\displaystyle\SumK\frac{\pK}{\ZK (\betaK )}   	\SumiK \exp\big(-\betaK\eiK\big)\,\PiK} \label{eq:exp_rho_hes2} 
\, .
\end{align}
Note that if the $\HK $'s (and hence the $\PK $'s) are time independent, then the time evolution $\rho(t)$ of the state is parametrized by only $2M$ variables, namely, either $\pK (t)$ and $\betaK (t)$ (Eq.~\eqref{eq:rhohypo}) or, equivalently, $\alphaK (t)$ and $\betaK (t)$ (Eq.~\eqref{eq:rhohypo2}). This is an important simplification that, as already referenced in Section~\ref{sec:intro}, when combined with the SEAQT dynamical equations has enabled the successful modeling of a broad range of mesoscopic systems. As shown below in Section~\ref{sec:rcce}, the HE approximation is equivalent to the translation into the nonequilibrium quantum thermodynamic framework of the constrained equilibrium approximation in the chemical kinetics framework and the quasi-equilibrium approximation in the dynamical systems framework.

The HE family in Eq.~\eqref{eq:alphaK} is now parametrized by $\{\pK,\betaK\}_{\smallK=1}^M$ subject to $\sumK\pK=1$ or, equivalently, by $\{\alphaK,\betaK\}_{\smallK=1}^M$  subject to the constraint $\sumK \ZK (\betaK ) \exp(-\alphaK)=1$ induced by normalization. Hence, the HE manifold has dimension $2M-1$.

Taking the logarithms of the HESS density operators (Eq.~\eqref{eq:rhoKeq}) and using   Eq.~\eqref{eq:exp_rho_hes} yields for $\ln\rho^\text{HE}$  a remarkably simple block-diagonal form where each sector  $\Hil_{\smallK}$ contributes an effective Hamiltonian term $\betaK \HK$ shifted by the sector-specific normalization $\alphaK \PK$. As a result the following relations hold for each sector:
\begin{align}
 \SKtilde^\text{HE}&=-\Boltz   \ln \rhoKtilde^\text{HE}  =\Boltz\betaK  \HK  + (\ln \ZK )\Boltz\PK \, , \label{eq:lnrhoKtilde}
\\
	\SK^\text{HE} &=\sK \PK +\SKtilde^\text{HE}=\Boltz\alphaK  \PK +\Boltz \betaK  \HK \, , \label{eq:SKHES}
\\
	\langle \SK^\text{HE} \rangle &=\Boltz\alphaK   \pK +\langle  \HK \rangle \Boltz \betaK \, , \label{eq:SKHESmean}
	\\
	\langle \SK^\text{HE}\HK \rangle &=\Boltz\alphaK   \langle  \HK \rangle + \Boltz \betaK \,\langle  \HK\HK \rangle\, , \label{eq:SKHKHESmean}
\end{align}
and for the overall system,
\begin{align}
   S^\text{HE}&=-\Boltz \ln\rho^\text{HE}=\Boltz\SumK(\alphaK \PK   +\betaK  \HK ) \, , \label{eq:lnrhoHES}
\\
	\langle S^\text{HE} \rangle& =\Boltz\SumK(\alphaK   \pK +\langle  \HK \rangle  \betaK )\, , \label{eq:lnrhoHESmean}
\end{align}
so that, for  arbitrary parameters $\alpha$ and $\beta$, the following ``nonequilibrium Massieu operator'' can be defined:
\begin{align} 
	M^\text{HE} &=S^\text{HE}-\Boltz\alpha I-\Boltz\beta H  = \SumK (  \SK^\text{HE} -\Boltz\alpha\PK-\Boltz \beta \HK  ) \label{eq:Massieu1}
\\
	\langle  M^\text{HE}\rangle& = \SumK (   \langle \SK^\text{HE} \rangle -\Boltz\alpha\pK  - \langle \HK \rangle\Boltz \beta  )\, .
\end{align}

Furthermore, following~\cite{Li2016a,Li2016b}, it is important to observe that Eq.~\eqref{eq:piKeq} implies
\begin{equation}  \label{eq:siKHES}
    \siK =-\Boltz\ln \piK =\sK+\Boltz\ln \ZK  +\Boltz\betaK \eiK =\Boltz\alphaK  +\Boltz\betaK \eiK 
\end{equation}
and, therefore, rewriting it for the index $j$ and subtracting and dividing by $\eiK -\ejK$ yields the  relations
\begin{equation}\label{eq:sijovereij}
    \frac{\siK -\sjK}{\eiK -\ejK}=\Boltz\betaK  \qquad \forall i_\smallK,j_\smallK\in \{1,\dots,M_K\}
\end{equation}
expressing a fingerprint feature of stable equilibrium, which  here (by Assumption HE5) holds only locally  within each HESS but not globally. For a global HES $\rho^\text{HE}$ to approach a stable equilibrium state $\rho^\text{SE}=\exp(-\beta^\text{SE} H)/\Tr[\exp(-\beta^\text{SE} H)]$, all the inverse temperatures $\betaK$ must converge to a common value   $\beta^\text{SE}$, all $\alphaK$'s converge to the common value   $\alpha^\text{SE}=\ln\sumK\ZK(\beta^\text{SE})$, and the $\pK$'s converge to $\pKSE=\ZK(\beta^\text{SE})/\sumL\ZL(\beta^\text{SE})$.

If a HE sector contains only a single distinct energy eigenlevel so that $M_\smallK =1$, then Eq.~\eqref{eq:piKeq} is satisfied identically for any $\betaK$. Such sectors carry population information but no meaningful internal temperature. In applications, partitions with $M_\smallK \geq 2$ for all sectors are typically chosen and, thus, carry a nontrivial internal equilibration structure.

When no decomposition is assumed, i.e., the whole Hilbert space is viewed as a single sector so that $\MK=1$, then assumption HE5  corresponds to a locally stable equilibrium state. In this trivial case, $\smallK$ only takes the value 1, $p_1=1$, $s_1=0$, $P_1=I$, $H_1=H$, and
\begin{equation}
    \langle S^\text{HE} \rangle =\Boltz\alpha_1   +\langle  H \rangle \Boltz \beta_1 \,,\qquad \alpha_1=\ln Z_1(\beta_1)\,,\qquad \dv{\alpha_1}{t}=-\langle H\rangle \dv{\beta_1}{t}\,.
\end{equation}

 \section{HE Assumptions for a Structured Composite of Subsystems}\label{sec:CSHE}
 
    When dealing with a composite system with overall  density operator $\rho$ on the system's Hilbert space $\Hil=\bigotimes_{\J=1}^{\M}\Hil_\J$, the composite--system hypoequilibrium (CSHE) approximation results when the following assumptions hold: 
    \begin{enumerate}
        [label=\textbf{(CSHE\arabic*):},start=1, labelsep=4pt, leftmargin=*]
       	  \item The subsystems are 
       	noninteracting, i.e.,
       	\begin{equation}
       		H=\sum_{\J=1}^{\M} H_\J\otimes I_\Jbar \,,
       	\end{equation}
       and in an uncorrelated state, i.e.,
       \begin{equation}
        	\rho=\bigotimes_{\J=1}^{\M} \rho_\J = \rho_\J\otimes\rho_\Jbar \,,
       \end{equation} 
       such that $\rho$ commutes with $H$, i.e.,
        \begin{equation}
        	\comm*{H}{\rho}=0 \qquad\text{ which implies } \comm*{H_\J}{\rho_\J}=0\ \ \forall \J \,.
       \end{equation} 
       Here, the subscript $\Jbar$ denotes the complement of subsystem $\J$ with Hilbert space
        \begin{equation}
    	   \Hil_\Jbar =\bigotimes_{\substack{\mathbb{L}=1\\ \mathbb{L}\ne\J}}^{\M}\, \Hil_\mathbb{L}  \,.
        \end{equation}
\item Each $\rho_\J$  is full rank, i.e., has no zero eigenvalues. 
\item Each $\rho_\J$ gives equal probability to each of the $g^\J_i$ corresponding eigenstates of each degenerate eigenvalue $\varepsilon^\J_{i}$ of $H_\J$, so that $\rho_\J$ and $H_\J$ share the same set of eigenprojectors $\PiJ$.
     \item Each subsystem's Hilbert space $\Hil_\J$  is decomposed into $M_\J$ spectral sectors, associated with the spectral decomposition of the local Hamiltonian operator $H_\J$, i.e.,
     	\begin{equation}
      H_\J        = 	\sumNJ  \eiJ\PiJ = \sumKJ \HKJ \quad\text{ with } \HKJ =	\sumiKJ \eiKJ\,\PiKJ\, , 
     	\end{equation}
based on an arbitrary partition of the set of $N_\J$  eigenvalues of $\rho_\J$ into  $M_\J$ disjoint subsets. The overall system's Hilbert space is, therefore, decomposed as 
     	\begin{equation}\label{eq:Hil_CS}
     		\Hil=\nolinebreak \bigotimes_{\J=1}^{\M}\bigoplus_{\smallK_\J=1}^{M_\J}\Hil_{\smallK_\J}^\smallbbJ   	\,
     	\end{equation}
     and resolutions of the identities $\PKJ$ within each subspace $\Hil_{\smallK_\J}^\smallbbJ $ in terms of the  mutually orthogonal eigenprojectors are given by
        	\begin{equation}\label{eq:identityK_CS}
     \PKJ=\SumiKJ  \PiKJ    	\,.
     \end{equation}
\item  A state $\rho$ is a ``CSHE state'' with respect to the chosen decomposition \eqref{eq:Hil_CS}  if, for each HESS $K_\J$, there exists an inverse temperature $\beta_{\smallK_\J}^\smallbbJ$ such that
   	\begin{equation}\label{eq:rhoKeqCS}
   	\rhoKJtildeHE = \frac{\PKJ \exp(-\betaKJ  H_\J   )\PKJ}{\ZK (\betaKJ )}, \qquad  \ZK (\betaKJ )=\Tr[\PKJ\exp(-\betaKJ  H_\J  )\PKJ]\, .
   	\end{equation}
   Following the same steps used to prove Eqs.~\eqref{eq:exp_rho_hes} and \eqref{eq:exp_rho_hes2}, it follows that  
   \begin{align}
   	\rho_\J^\text{HE}= \SumKJ \pKJ \,\rhoKJtildeHE  &=
   \SumKJ 	\SumiKJ \exp\big(-\alphaKJ-\betaKJ\eiKJ\big)\,\PiKJ \label{eq:exp_rho_CSHE}  \\ &= \SumKJ  \frac{\pKJ}{\ZK (\betaKJ )}	\SumiKJ \exp\big(-\betaKJ\eiKJ\big)\,\PiKJ \label{eq:exp_rho_CSHE2} \, .
   \end{align}
   Equivalently, $\rho^\text{HE}=\bigotimes_{\J=1}^{\M} \rho^\text{HE}_\J$ where
      	\begin{equation}\label{eq:piKJeqCS}
      \rho_\J^\text{HE}=  \SumKJ 	\SumiKJ \piKJ \,\PiKJ\quad  \text{ with }	\piKJ=\frac{\pKJ}{\ZK (\betaKJ )}\exp\big(-\betaKJ\eiKJ\big)\, .
   \end{equation}
   
\end{enumerate}

Following the same procedure as in Section \ref{sec:hes}, the subsystems' nonequilibrium Massieu operators for arbitrary parameters $\alpha_\J$ and $\beta_\J$ (that will take explicit values in Appendix \ref{sec:seaqt}) may be written as
\begin{align} 
	M_\J^\text{HE} &=S_\J^\text{HE} -\Boltz\alpha_\J I_\J-\Boltz\beta_\J H_\J  = \SumKJ (  \SKJHE-\Boltz\alpha_\J \PKJ -\Boltz \beta_\J \HKJ  )\nonumber  \\ &=\Boltz \SumKJ [ (\alphaKJ- \alpha_\J)  \PKJ +(\betaKJ- \beta_\J) \HKJ ] \label{eq:MassieuJ}
\end{align} 
where $S_\J^\text{HE}=-\Boltz\ln\rho_\J^\text{HE}$ and
\begin{align} 
\SKJHE &=\sKJ \PKJ +\SKJtildeHE=\Boltz\alphaKJ  \PKJ +\Boltz \betaKJ  \HKJ\,, \\
\SKJtildeHE&=-\Boltz   \ln \rhoKJtildeHE  =\Boltz\betaKJ  \HKJ  + (\ln \ZKJ )\Boltz\PKJ\,, \\
\sKJ&=-\Boltz\ln  \pKJ \,.
\end{align}

\section{Consistency of the HE approximation with the RCCE approach}\label{sec:rcce}

In this section, the  HE approximation is shown to fit precisely within the framework of the RCCE method for model reduction. This method was introduced and applied by Keck and coworkers~\cite{Keck1971,Keck1979,BerettaKeck1986,Keck1990,Beretta2012}  as a thermodynamically consistent method for obtaining accurate results in combustion modeling applications that involve complex chemical kinetic schemes. Many others have worked on variations, improvements, generalizations, and geometrizations of this model reduction technique~\cite{Pope2006,Gorban2006,Chiavazzo2012,Hadi2015,Beretta2016,Karlin2016,Rivadossi2018}.  

In the present HE framework, it is assumed that the dynamics has bottlenecks  associated with the flow of probability and energy between different HESS's so that the irreversible redistribution of  probabilities $\pK $ and  partial energies $\langle \HK \rangle$ among different HESS's is much slower than the probability redistribution within each HESS. As a result of the rapid equilibration within each HESS,  each density operator $\rhoKtilde $ rapidly relaxes towards the maximum \textit{proper} entropy $\langle \SK \rangle_\smallK$ compatible with the local normalization condition $\Tr(\rhoKtilde)  = 1$ and the current value of the \textit{proper} mean energy  $\langle \HK \rangle_\smallK$, namely, towards  the $\rhoKtilde $ given by Eq.~\eqref{eq:rhoKeq}. Figure 1 provides a pictorial representation of the HE approximation with the help of the energy-vs-entropy diagrams developed in \cite{GB1991,Beretta2026}.

\begin{figure}[!htbp]
    \includegraphics[width=\textwidth]{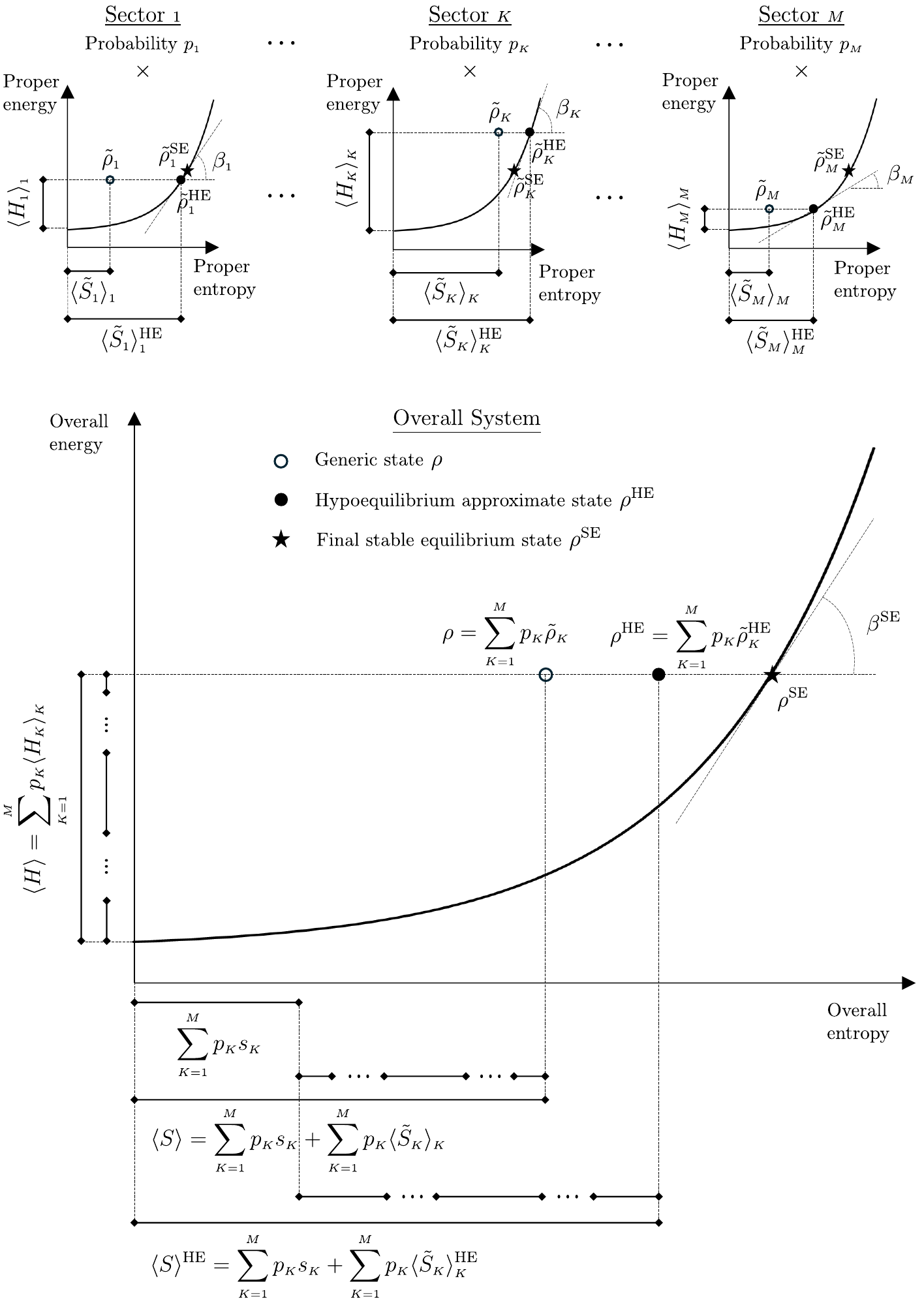}
    \caption{\label{fig:Figure1}
    Top: representation of the HESS density operator $\tilde\rho_\smallK$ and its HE approximation $\tilde\rho_\smallK^\text{HE}$ on the proper-energy--vs--proper-entropy diagram of each sector $\smallK$. Bottom: representation of the overall density operator $\rho$ and its HE approximation $\rho^\text{HE}$ on the energy-vs-entropy diagram of the overall system.  The time evolution is assumed to redistribute probabilities much more rapidly within each HE sector than among different sectors. Therefore, each  nonequilibrium HESS  state $\tilde\rho_\smallK$ approaches rapidly the sector maximal-proper-entropy state $\tilde\rho_\smallK^\text{HE}$ with the same mean proper  energy $\langle \HK\rangle_\smallK$. The corresponding overall states $\rho$ and $\rho^\text{HE}$ are constructed on the overall energy--vs--entropy diagram using the additivity relations summarized in Table 1. The stars denote the final stable equilibrium state when all the $\betaK$'s reach the same value $\beta^\text{SE}$.}
\end{figure}

In terms of RCCE terminology, it is assumed that the relatively slow, rate-controlling constraints associated with the bottlenecks of the overall dynamics are the spectral-sector total probabilities $\pK $ and partial mean energies $\langle \HK \rangle$. Therefore, at any given time during the evolution, the state $\rho$ is assumed to be well approximated by the state that maximizes the overall entropy $\langle S\rangle=-\Boltz\Tr(\rho\ln\rho)$ subject to these constraints. 
The constrained maximization can be written as
\begin{equation}
    \max_{\rho}|_{\pK ,\langle \HK \rangle, \PK, \HK }\ \ -\Boltz\Tr(\rho\ln\rho) \qquad  \mbox{ subject to } \Tr(\rho\PK) =\pK \, \mbox{  and } \Tr(\rho \HK) =\langle \HK \rangle\, .
\end{equation}
Introducing the Lagrange multipliers $\alphaK -1 $ and $\betaK $ (in the RCCE approach $\alphaK  $ and $\betaK $ are called ``constraint potentials'') and using Eqs.~\eqref{eq:pK}, \eqref{eq:EK}, and~\eqref{eq:sk_avg}, the equivalent unconstrained maximization becomes 
 \begin{equation}
\max_{\piK}|_{\giK ,\eiK }\ \ -\SumK\SumiK\piK\giK\ln\piK - \SumK(\alphaK -1) \SumiK\piK\giK  -\SumK \betaK \SumiK\piK\giK\eiK\, .
\end{equation}
The solution is readily found to be
 \begin{equation}\label{eq:RCCE1}
    \piK = \exp( -\alphaK  - \betaK   \eiK)\, .
\end{equation}
Substituting this into the constraints yields the Lagrange multipliers in terms of $\pK $ and $\langle \HK \rangle$.  From the first constraint, 
\begin{equation}\label{eq:constr1}
    \pK =\SumiK\giK \exp( -\alphaK  - \betaK   \eiK) = \exp(-\alphaK ) \Tr[\exp(-\betaK  \HK  )]  = \exp(-\alphaK ) \ZK (\betaK )\, ,
\end{equation}
Eq.~\eqref{eq:alphaK}, i.e., $\exp(-\alphaK )=\pK/\ZK (\betaK )$ where $\ZK (\betaK )=\Tr[\exp(-\betaK  \HK  )]$ follows so that Eq.~\eqref{eq:RCCE1} can be rewritten as
\begin{equation}
 	\piK=\pK 
 	\frac{\exp(-\betaK  \eiK  )}{\ZK (\betaK )}\, ,
\end{equation}
and, therefore,
\begin{equation}
    \rho^\text{RCCE}=\SumK\pK 
    \frac{\PK\exp(-\betaK  \HK  )\PK}{\ZK (\betaK )}=\SumK\pK\,\rhoKtilde^\text{HE} ,
\end{equation}
which  coincides with Eq.~\eqref{eq:rhohypo}.  Substituting into the second constraint yields the relation
\begin{equation}\label{eq:constr2}
    \langle \HK \rangle=\frac{ \pK }{\ZK (\betaK )}\Tr [\HK \exp(-\betaK  \HK  )] \, ,
\end{equation}
which can be solved to obtain $\betaK  =\betaK (\langle \HK \rangle/\pK )$ and together with Eq.~\eqref{eq:constr1} yields $\alphaK  =\alphaK (\pK ,\langle \HK \rangle/\pK )$. 

Recalling that $\langle \HK \rangle = \langle \HK \rangle_\smallK\,\pK   $   (Eq.~\eqref{eq:EKK}), these relations can be rewritten in terms of the \textit{proper} mean HESS energy instead of the  \textit{local} mean HESS energy since $\betaK  =\betaK (\langle \HK \rangle_\smallK )$ and $\alphaK  =\alphaK (\pK ,\langle \HK \rangle_\smallK )$.  Within the RCCE method, adopting the constraint potentials (here $\alphaK$ and $\betaK$) as the independent variables~\cite{Beretta2012} avoids having to solve the above system of equations to obtain the values of $\alphaK$ and $\betaK$ from the current values of $\pK$ and $\langle \HK \rangle$ at every time step during the integration of the evolution equation. 

Finally, the RCCE description of nonequilibrium states (and hence also the HE description) follows  the same logic as the standard model for chemically reacting systems where nonequilibrium states are assigned the properties of the stable equilibrium state of a ``surrogate'' system (see~\cite{GB1991,Gyftopoulos2015}) with the same composition, volume, and energy, but with all reactions frozen. This amounts to treating the chemical reactions as the bottlenecks of the dynamics, the corresponding set of constraints being the set of all species amounts.The success of the RCCE method, thus, hinges on the ability to identify the bottleneck constraints. Equivalently, in the present SEAQT context, the key to successful model reduction is the ability to identify the $M$ subsets of energy eigenvalues subject to rapid irreversible redistribution within each HESS so that the focus can be on the slow (bottleneck) dynamics that controls the energy and entropy exchanges between the different HESS's.

\section{HE--SEAQT for an Unstructured and Isolated System}\label{sec:heseaqt_single}

Appendix \ref{sec:seaqt} provides a review of the foundational assumptions of the original SEAQT 
formalism for a general system with internal structure, while Appendix \ref{sec:heseaqt} merges these assumptions with the HE assumptions discussed in Section \ref{sec:CSHE} and completes them with two additional assumptions to obtain the HE--SEAQT formulation for a system with noninteracting subsystems. Under this set of HE--SEAQT assumptions, there is no interaction Hamiltonian and each subsystem evolves independently of the others. It, therefore, suffices to consider a single, unstructured  system that can be modeled without  internal subdivision into separated, noninteracting, and uncorrelated subsystems. The HE--SEAQT equation of motion takes the same form as Eqs.~(\ref{eq:SEA_generalHE}--\ref{eq:multipliers_generalHEAB2}) but without the $\mathbb{J}$ super- and subscripts, i.e.,
\begin{equation}\label{eq:SEA_singleHE} 
	\dv{\rho}{t}=-\acomm*{\mathcal{D}_\rho}{\rho}=\SumK \frac{\pK}{\tauK} [  (\alphaK- \alpha)    \rhoKtilde +(\betaK- \beta) \HK \rhoKtilde ]\,,
\end{equation}
\begin{equation}\label{eq:SEA_singleHES} 
	\dv{S}{t}=-\Boltz\SumK \frac{1}{\tauK} [  (\alphaK- \alpha)    \PK +(\betaK- \beta) \HK  ] \,,
\end{equation}
and the SEA nonequilibrium potentials $\alpha$ and $ \beta$ are  functions of all the constraint potentials $\alphaK$ and $\betaK$ which may be written as
\begin{align}\label{eq:multipliers_singleHE1} 
	\Boltz\alpha &= \frac{B_{S} B_{HH}-B_{H} B_{SH} }{ B_{HH}-B_{H} B_{H}}=\langle s \rangle_w - \Boltz\beta\,\langle \varepsilon  \rangle_w \,,\\
	\Boltz\beta &= \frac{ B_{SH}   - B_{S} B_{H}}{ B_{HH} -B_{H} B_{H}}= \frac{
		\langle \Delta_w s\Delta_w\varepsilon \rangle_w}{\langle \Delta_w \varepsilon\Delta_w\varepsilon \rangle_w}\,,\label{eq:multipliers_singleHE2}
\end{align}
where
{\allowdisplaybreaks    
\begin{align}\label{eq:multipliers_singleHEB1} 
	B_{H} &= \Tau \SumK \frac{\langle \HK \rangle }{\tauK}= \Tau\SumK \frac{\pK }{\tauK}\langle \HK \rangle_\smallK=\SumK \SumiK \wiK\eiK=\langle \varepsilon \rangle_w\,,\\
	B_{S} & = \Tau\SumK \frac{\langle \SK \rangle }{\tauK}=\Boltz\Tau \SumK\frac{\pK }{\tauK}\big(\alphaK+\betaK \langle \HK \rangle_\smallK\big) =\SumK \SumiK \wiK\siK=\langle s \rangle_w\,, \\
	B_{HH} &= \Tau \SumK \frac{\langle \HK\HK \rangle }{\tauK}=\Tau\SumK \frac{\pK }{\tauK}\langle \HK\HK \rangle_\smallK=\SumK \SumiK \wiK(\eiK)^2=\langle \varepsilon^2 \rangle_w\,,\\
	B_{SH} &= \Tau\SumK \frac{\langle \SK\HK \rangle }{\tauK}= \Boltz\Tau\SumK\frac{\pK }{\tauK}\big(\alphaK \langle \HK \rangle_\smallK+\betaK \langle \HK\HK \rangle_\smallK\big)=\SumK \SumiK \wiK\eiK\siK=\langle s\varepsilon \rangle_w \,. \label{eq:multipliers_singleHEB2}
\end{align}
Here, $\Tau$, $\langle x \rangle_w$, and $\langle \Delta_w x\Delta_w y \rangle_w$ denote overall weighted averages and covariances with respect to the dimensionless weights $\wiK$ and the weighted deviations (from the overall weighted average) $\Delta_w \xiK= \xiK-\langle x \rangle_w$ defined as follows
\begin{align}
	\wiK&=\Tau\frac{\piK\giK}{\tauK} \,,& \frac{1}{\Tau}=\SumK \frac{\pK}{\tauK}&=\SumK \SumiK \frac{\piK\giK}{\tauK}\,,\\
	\langle x \rangle_w &= \SumK \SumiK \wiK\xiK \,,& \langle \Delta_w x\Delta_w y \rangle_w &= \SumK \SumiK\wiK\Delta_w \xiK\Delta_w \yiK \label{eq:multipliers_singleHEB3}
\end{align}

Denoting by $\miK$ the eigenvalues of the local nonequilibrium Massieu operator of sector $\smallK$, ${\MK}$, and recalling Eqs.~\eqref{eq:pikHE} and~\eqref{eq:sikHE},  the following  relations corresponding to normalization, mean-energy conservation, and entropy production rate are readily verified:
\begin{align}
\piK&=\exp(-\alphaK -\betaK  \eiK)  \,,\qquad\qquad\quad 
\siK=\Boltz\alphaK +\Boltz\betaK  \eiK  \,,\\
\MK&=\SK-\Boltz\alpha\PK-\Boltz\beta\HK \,,\qquad\qquad	\miK=	\siK-\Boltz\alpha -\Boltz\beta\,  \eiK   \,,\\
	\langle m \rangle_w&=\langle\Delta_w m \rangle_w=	\SumK \SumiK \wiK\miK = 0\,,\\	
	\langle m\varepsilon \rangle_w&=\langle \Delta_w m\Delta_w \varepsilon \rangle_w=\SumK \SumiK \wiK\miK\eiK = 0\,,\\
	\Boltz\Tau\dv{\langle S\rangle}{t}&= \SumK \SumiK \wiK(\miK)^2 = \langle m^2 \rangle_w= \langle \Delta_w m\Delta_w m \rangle_w \ge 0\,.
\end{align}
Note that the overall nonequilibrium Massieu operator $M=\sumK \MK$ has zero weighted mean value but a nonzero weighted variance proportional to the overall entropy production rate, which vanishes only at stable equilibrium states.
}

Recalling that $\piK\giK=\Tr(\rho\PiK)$, $\siK=-\Boltz\ln\piK$, $\pK=\Tr(\rho\PK)$, $\langle \HK \rangle=\langle \HK \rangle_\smallK\,\pK $, and $\sK=-\Boltz\ln\pK$, Eq.~\eqref{eq:SEA_singleHE} yields the following relations:
\begin{equation}\label{eq:pikdotHES}
	\tauK\ddt{\piK }=  (\alphaK -\alpha ) \piK +(\betaK -\beta)\,\eiK \piK  \,,
\end{equation}
\begin{equation}\label{eq:sikdotHES}
		\frac{\tauK}{\Boltz}\ddt{\siK }=  (\alpha-\alphaK ) +(\beta- \betaK )\,\eiK   \,,
\end{equation}
\begin{equation}\label{eq:pkdotHES}
	\tauK\ddt{\pK }=  (\alphaK -\alpha ) \pK +(\betaK -\beta) \langle \HK \rangle = (\alphaK -\alpha ) \pK +(\betaK -\beta) \langle \HK \rangle_\smallK\,\pK  \,,
\end{equation}
\begin{equation}\label{eq:skdotHES}
	\frac{\tauK}{\Boltz}\ddt{\sK}=  (\alpha-\alphaK )+(\beta- \betaK ) \frac{\langle \HK \rangle}{\pK }= (\alpha-\alphaK )+(\beta- \betaK ) \langle \HK \rangle_\smallK \,,
\end{equation}
\begin{equation}\label{eq:sumpkconst}
	\SumK \frac{1}{\tauK}  [(\alphaK -\alpha ) \pK +(\betaK -\beta) \langle \HK \rangle] = 0  \,,
\end{equation}
\begin{equation}\label{eq:energyconst}
	\SumK \frac{1}{\tauK}  [(\alphaK -\alpha ) \langle \HK \rangle +(\betaK -\beta) \langle \HK\HK \rangle] = 0  \,,
\end{equation}
\begin{equation}\label{eq:entropy_production_SEAHES} 
	\dv{\langle S\rangle}{t}=\Boltz\SumK\frac{1}{\tauK}\Tr\left(\rhoKtilde\,\left[  (\alphaK- \alpha)    \PK +(\betaK- \beta) \HK  \right]^2\right) \,,
\end{equation}
where $\expval{\HK\HK}=\Tr(\rho\HK\HK)=\pK\,\Tr(\rhoKtilde\HK\HK)=\pK\,\langle\HK\HK\rangle_\smallK$ and the systems of Eqs.~\eqref{eq:sumpkconst} and~\eqref{eq:energyconst} for normalization and the conservation of mean energy are entirely equivalent to the system of Eqs.~\eqref{eq:multipliers_singleHE1} and~\eqref{eq:multipliers_singleHE2}.

It is evident from Eq.~\eqref{eq:pkdotHES} that if a $\pK $ is zero at one time then it must be zero at all times. In other words, an unpopulated HE sector remains unpopulated under SEA dynamics, a feature that extends to the HE framework the known feature of SEA dynamics~\cite{Beretta1984} whereby zero eigenvalues of the density operator remain zero and nonzero ones remain nonzero (although they can approach zero much like $e^{-t}\rightarrow 0$ as $t\rightarrow \infty$). This feature holds also for Eq.~\eqref{eq:pikdotHES} and implies that if the $\piK$'s start all positive, as is required for the density operator to be strictly positive, they can never cross zero, whether the equation of motion is evolved forward or backward in time. Since the right-hand side of Eq.~\eqref{eq:entropy_production_SEAHES} is non-negative, it follows that under HE--SEAQT the rate of change of the overall system's entropy is non-negative in forward time and non-positive in backward time.

Subtracting from Eq.~\eqref{eq:sikdotHES} the same equation written for another energy eigenlevel $\ejK$ in the same sector $\smallK$, and recalling Eq.~\eqref{eq:pikHE}, yields
\begin{equation}\label{eq:inv_manifold}
	\frac{\tauK}{\Boltz}\ddt{(\siK -\sjK)}= (\betaK -\beta)\,(\eiK -\ejK)= \tauK\ddt{(\alphaK+\betaK\eiK -\alphaK-\betaK\ejK)}  \,,
\end{equation}
which reduces to
\begin{equation}\label{eq:betaKdot}
	\tauK\ddt{\betaK}= \beta-\betaK  \,.
\end{equation}
Similarly, dividing Eq.~\eqref{eq:sikdotHES} by $\eiK$, subtracting the resulting equation written for another energy eigenlevel $\ejK$ in the same sector $\smallK$, and again using Eq.~\eqref{eq:pikHE} yields
\begin{equation}
	\frac{\tauK}{\Boltz}\ddt{(\siK/\eiK -\sjK/\ejK)}= (\alphaK -\alpha )\,\left(\frac{1}{\eiK} -\frac{1}{\ejK}\right)= \tauK\dv{(\alphaK/\eiK+\betaK -\alphaK/\ejK-\betaK)}{t}  \,,
\end{equation}
which reduces to
\begin{equation}\label{eq:alphaKdot}
	\tauK\ddt{\alphaK}= \alpha-\alphaK  \,.
\end{equation}
Note that the overall energy spectrum can be shifted by a constant for Hamiltonians with zero eigenenergies so that the new $\eiK$'s are nonzero.
Together with Eqs.~(\ref{eq:multipliers_singleHE1}--\ref{eq:multipliers_singleHEB2})---which give $\alpha$ and $\beta$  in terms of the $\alphaK$'s and $\betaK$'s---Eqs.~\eqref{eq:betaKdot} and \eqref{eq:alphaKdot} form a differential-algebraic system of nonlinearly coupled equations that automatically satisfies the normalization and energy conservation constraints. For a given set of initial values  $\{\alphaK(0),\betaK(0),{\scriptstyle K{=}1},{\scriptstyle \dots},{\scriptstyle M}\}$,  which identify an initial HES, this system can be readily solved numerically to yield the time evolution of the  $2M$ HE constraint potentials  $\alphaK(t)$ and $\betaK(t)$, along with that of the overall system's SEA nonequilibrium potentials $\alpha(t)$, $\beta(t)$. It is clear from Eqs.~\eqref{eq:betaKdot} and \eqref{eq:alphaKdot}  that the time evolution does not end until (i) $\beta$ and all the $\betaK $'s converge to the same value, $\beta(\infty)=\beta^\text{SE}$, and  (ii) $\alpha$ and all the $\alphaK $'s converge to the same value, $\alpha(\infty)=\alpha^\text{SE}$. This observation corroborates the interpretation of $\beta$ as playing the role of a dynamic nonequilibrium inverse temperature.

As the state evolves toward the  canonical Gibbs state $\rho^\text{SE}=\exp(-\beta^\text{SE}H)/Z(\beta^\text{SE})$ with inverse temperature $\beta^\text{SE}$ identified by the initial mean energy $\Tr(\rho(0) H)$, the values of the SEA potentials $\alpha$ and $ \beta$ keep changing as the  local energies $\langle \HK\rangle$ are redistributed among the HE sectors.

Substituting Eqs.~\eqref{eq:betaKdot} and \eqref{eq:alphaKdot} into  \eqref{eq:sikdotHES} yields
\begin{equation}
	\frac{1}{\Boltz}	\dv{\siK}{t}=\dv{\alphaK}{t} +\dv{\betaK}{t}  \eiK  \, .
\end{equation}
which coincides with Eq.~\eqref{eq:sikdotHE} and, therefore, proves that the HE manifold is  invariant under the  SEAQT equation of motion. It is in fact a strongly invariant manifold, because, as noted above, every density operator in $\PHES$ belongs to a trajectory lying entirely within $\PHES$ and well-defined in the interval  $ -\infty\le t\le  \infty$, evolving in forward time from a minimum value to a maximum value of the entropy. As emphasized and exemplified in \cite{Beretta2016}, this feature of SEAQT, which extends to HE--SEAQT, can be interpreted as implementing a strong version of the principle of causality, whereby knowing the state at time $t{\,=\,}0$ identifies a unique trajectory in state space covering the future as well as the past. Mathematically, the time evolution is governed by a temporally reversible one-parameter group (not a semigroup as in the GKSL equation), which nevertheless describes a thermodynamically irreversible evolution, establishing as dynamical theorems both the principle of entropy non-decrease in forward time and the Hatsopoulos-Keenan statement of the second law, i.e., the conditional stability of the maximum entropy (Gibbs) states (`conditional' in the technical sense detailed in \cite{Beretta1986}).

Now, recalling Eqs.~\eqref{eq:SKHESmean} and \eqref{eq:SKHKHESmean} and using Eqs.~\eqref{eq:betaKdot} and \eqref{eq:alphaKdot}, the rates of change of the sector energies and entropies during the irreversible redistribution of populations may be computed from the relations
\begin{equation}\label{eq:Edot1}
		\dv{\langle \HK \rangle}{t}=-\expval{\HK}\dv{\alphaK}{t} -  \expval{\HK\HK} \dv{\betaK}{t}\, ,
\end{equation}
\begin{equation}\label{eq:Sdot1}
\frac{1}{\Boltz}	\dv{\langle \SK\rangle}{t}= \big(\pK-\expval{\SK}/\Boltz\big)\dv{\alphaK}{t}+  \big(\expval{\HK}-\expval{\SK\HK}/\Boltz\big)\dv{\betaK}{t}\, .
\end{equation}
Furthermore, it is  worth emphasizing another important feature of the HE (and RCCE) approximation, namely, that even though it reduces the description of the dynamics from the full set of differential equations for the $N^2-1$ real parameters needed to determine a state operator $\rho$ to the possibly much smaller set of $2M$ constraint potentials $\alphaK$ and $\betaK$, it does not reduce the state description. At any instant of time, via Eq.~\eqref{eq:pikHE},  the values of the $\alphaK$'s and $\betaK$'s determine the full density operator and, therefore, allow computation of the mean values of all the system's properties.

The SEA potentials $\alpha$ and $\beta$ can be interpreted as the effective nonequilibrium properties that mediate the relaxation of the entire system. Their final values emerge from a competitive consensus between sectors, where each sector $\smallK$ exerts a ``leverage'' proportional to its statistical weight $\pK$, its internal energy fluctuations, and its relaxation speed $1/\tauK$. To formalize this, the numerators and denominators of Eqs.~\eqref{eq:multipliers_singleHE2} are rewritten as follows:
\begin{equation}
    \beta = \frac{\SumK  [ \langle \varepsilon^2 \rangle_w^\smallK  - (\langle \varepsilon \rangle_w^\smallK)^2 ]\,w_\smallK \betaK }{\langle \varepsilon^2 \rangle_w - (\langle \varepsilon \rangle_w)^2} + \frac{\SumK  (\langle s \rangle_w^\smallK - \langle s \rangle_w)(\langle \varepsilon \rangle_w^\smallK - \langle \varepsilon \rangle_w)\,w_\smallK}{\langle \varepsilon^2 \rangle_w - (\langle \varepsilon \rangle_w)^2} \,,
\end{equation}
where the following local weighted averages are defined:
\begin{equation}
	\langle x \rangle_w^\smallK = \frac{1}{w_\smallK}\SumiK \wiK\xiK\,,\qquad 
	w_\smallK = \SumiK \wiK \,.
\end{equation}
This decomposition reveals that the SEA global inverse temperature $\beta$ is shaped by two distinct physical contributions. The first term represents a weighted average of the HESS local $\beta_K$ values where the importance of each sector is scaled by its energy fluctuations. This implies that sectors with broader energy distributions (i.e., higher heat capacities) contribute more significantly to the instantaneous value of $\beta$ which, as shown by Eqs.~\eqref{eq:betaKdot} and \eqref{eq:alphaKdot}, acts as a common attractor for the local $\beta_K$ values. The second term accounts for the entropy--energy covariance between sector weighted averages. It captures how the displacement of a sector’s mean energy and mean entropy from the respective global averages affects the overall target slope of the entropy--energy relation.

Given these relations, a single sector $\smallL$ can dominate the global $\beta$ and impose its local value $\betaL$---thus acting as an internal heat bath---under three specific conditions: (i) Statistical Weight: when $w_\smallL \to 1$, meaning the sector encompasses nearly the entire system; (ii) Fluctuation Dominance: when the internal variance of sector $\smallL$ is much larger than that of all other sectors combined, effectively overwhelming their contributions and cross-couplings; (iii) Geometric Leverage: when sector $\smallL$ is located at an extreme energy mean ($\langle \varepsilon \rangle_w^\smallL \gg \langle \varepsilon \rangle_w$), acting as a ``leverage point'' that pivots the global regression line toward its own local parameters. In these cases, sector $\smallL$ acts as a stabilizer: its high ``nonequilibrium thermal inertia'' allows it to absorb significant energy fluctuations without shifting its own local potentials, effectively ``pinning'' the global $\alpha$ and $\beta$ to its local values. Consequently, the heat bath dictates the target temperature of the system, forcing smaller, more volatile sectors to align and equilibrate. In the quasihomogeneous near-equilibrium limit, where $\betaK \approx \beta$, the second term vanishes if the sector averages align perfectly along the global equilibrium line, effectively realizing a form of energy equipartition across the HE ensemble.

\section{NH-HE--SEAQT: Model of Non-Hamiltonian  Heat Interaction Between Unstructured Systems}\label{sec:heseaqt_heat}

Building on the sectoral decomposition developed for an isolated system in Section~\ref{sec:heseaqt_single}, the model is now extended to describe heat interactions between two or more systems.  
The suggestions in \cite{Li2016a,Li2018b} for a heuristic extension of the SEA and HE mathematical frameworks are followed to develop an effective and thermodynamically consistent model of heat interactions between systems. This approach is fundamentally different from that discussed in Section \ref{sec:heseaqt_single} and Appendices \ref{sec:seaqt} and \ref{sec:heseaqt}, where energy exchanges between subsystems can only occur via the effects of an interaction Hamiltonian $V$ through the von Neumann  (Hamiltonian) term in the equation of motion (Eqs.~\eqref{eq:SEA_Composite_simp} and \eqref{eq:SEA_local}). Here, instead, energy exchanges between subsystems are modeled via ``entropic coupling'' provided by a less constrained SEA dissipative term in the equation of motion.

All the HE assumptions discussed so far and the SEA assumptions reviewed in Appendices \ref{sec:seaqt} and \ref{sec:heseaqt} are adopted except for the following important modification. As detailed in Appendix \ref{sec:seaqt}, the variational principle that leads to  the composite-system version of the SEA equation of motion is stated as follows: 

\begin{enumerate}[label=\textbf{(SEAQT\arabic*):}, labelsep=4pt,start=3, leftmargin=*]\item
	The  dissipative part of the evolution equation ensures that, with respect to a local dissipative metric $\hat G_\J$, the direction of the local trajectory $\gamma_\J(t)$, maximizes the local contribution,  $\dot{s}|_\J$, to the overall system's entropy production rate. Under the constraints $\dot{c_q}|_\J=0$ which guarantee that the dissipative part of the dynamics does not contribute to the rates of change of the locally perceived  global charges $C_q$ (so that these emerge as constants of the motion if they are conserved also by the Hamiltonian part, i.e., if $\comm{H}{C_q}=0$).  \end{enumerate}
The rates of change of the overall system entropy, $\langle S\rangle$, and of the overall system mean value of $Q$ linear charges $\langle C_q\rangle=\Tr(\rho C_q)$, are written as
\begin{alignat}{2}
	\dv{\langle S\rangle}{t}&= \sum_{\J=1}^\M\dot{s}|_\J &\qquad \dot{s}|_\J&=\Big(2(S)^\J_\rho\gamma_\J\Big|\dot\gamma^{d}_\J\Big)\,, \\
	\dv{\langle C_q\rangle}{t}&= \sum_{\J=1}^\M\dot{c_q}|_\J&\qquad \dot{c_q}|_\J&=\Big(2(C_q)^\J_\rho\gamma_\J\Big|\dot\gamma^{d}_\J\Big)\,, 
\end{alignat}
exhibiting additive contributions from the $\M$ subsystems. Introducing the Lagrange multipliers $\vartheta_q^\J$ and $\tau_\J$ for the constraints, the SEAQT $\dot\gamma^d_\J$'s are found by solving the $\M$ local maximization problems
\begin{equation}\label{eq:variationalMain2} 
	\max_{\dot\gamma^{d}_\J}\ \Upsilon_\J= \dot{s}|_\J - \sum_{q=1}^Q\vartheta_q^\J\, \dot{c_q}|_\J -\frac{\Boltz\tau_\J}{2} \Big(\dot\gamma^d_\J\Big|\,\hat G_\J\,\Big|\dot\gamma^d_\J\Big)\,,\quad\text{ for every }\J=1,\dots,\M
\end{equation}
where the last constraint corresponds to the condition $(\dd \ell^d_\J/\dd  t)^2= \text{constant}$, necessary for maximizing with respect to direction only (see~\cite{Beretta2014,Gheorghiu2001a,Gheorghiu2001b} for more details). Since the local maximization problems \eqref{eq:variationalMain2} are independent, they can also be rewritten as a single, equivalent overall maximization problem, namely,
\begin{equation}
	\max_{\dot\gamma^{d}_\J}\ \Upsilon= \sum_{\J=1}^\M\dot{s}|_\J - \sum_{\J=1}^\M\sum_{q=1}^Q\vartheta_q^\J\, \dot{c_q}|_\J -\sum_{\J=1}^\M\frac{\Boltz\tau_\J}{2} \Big(\dot\gamma^d_\J\Big|\,\hat G_\J\,\Big|\dot\gamma^d_\J\Big)\,,
\end{equation}
 The first two charges are always the identity operator, $C_1\,{=}\,I$, which implements the $\Tr\rho\,{=}\,1$ constraint and the Hamiltonian operator, for $C_2\,{=}\,H$, which implements the $\Tr\rho H\,{=}\,\text{constant}$ constraint. The Lagrange multiplier $\vartheta_2^\J$ (usually renamed $\Boltz\beta_\J$) plays the role of ``local nonequilibrium inverse temperature'' conjugated with the locally perceived energy, and for the stable equilibrium states of the SEA dynamics, it coincides with the  thermodynamic inverse temperature.

The present heuristic model of heat interaction proposed in \cite{Li2016b}, instead, adopts the following modified assumption, which is called non-Hamiltonian (NH) here because it results in energy and entropy exchanges between subsystems that are  driven directly by the SEA dissipative term in the equation of motion. 

\begin{enumerate}[label=\textbf{(NH-SEAQT\arabic*):}, labelsep=4pt,start=3, leftmargin=*]\item
	The  dissipative part of the evolution equation  ensures that, with respect to a local dissipative metric $\hat G_\J$, the direction of the local trajectory $\gamma_\J(t)$, maximizes the local contribution,  $\dot{s}|_\J$, to the overall system's entropy production rate, under local conservation constraints $\dot{c_q}|_\J=0$ of the locally perceived  global charges $C^\text{local}_q$, for all $q$'s except $q=2$ corresponding to the Hamiltonian $C_2=H$, for which the conservation constraint is global (not local). [To model heat-and-diffusion interactions, the same exception is also extended in \cite{Li2018a} to the number-of-particle operator(s) $C_3=N$ ($C_{2+i}=N_i$).  \end{enumerate}
Therefore, the less-constrained overall maximization problem,
\begin{equation}\label{eq:variationalLess} 
	\max_{\dot\gamma^{d}_\J}\ \Upsilon= \sum_{\J=1}^\M\dot{s}|_\J - \sum_{\J=1}^\M\sum_{q\ne 2}^Q\vartheta_q^\J\, \dot{c_q}|_\J - \vartheta_2\sum_{\J=1}^\M\, \dot{c_2}|_\J  -\sum_{\J=1}^\M\frac{\Boltz\tau_\J}{2} \Big(\dot\gamma^d_\J\Big|\,\hat G_\J\,\Big|\dot\gamma^d_\J\Big)\,,
\end{equation}
is adopted so that the constraints of locally-perceived mean energy conservation within each subsystem are replaced by a single constraint of global mean energy conservation.

Setting the variational derivatives of $\Upsilon$ with respect to each $|\dot\gamma^{d}_\J)$ equal to zero, yields, in terms of the ``locally perceived  nonequilibrium Massieu operators'' $(M)^\J_\rho$,
\begin{equation} \label{eq:NHSEA_general}
\Big|\dot\gamma^d_\J\Big)=\frac{1}{\Boltz\tau_\J}\hat G_\J^{-1} 	\Big|2(M)^\J_\rho\gamma_\J\Big)\,, \qquad 
	(M)^\J_\rho=(S)^\J_\rho-\sum_{q\ne 2}^Q\vartheta_q^\J(C_q)^\J_\rho-\vartheta_2 (C_2)^\J_\rho \,.
\end{equation}
where the local Lagrange multipliers $\vartheta_q^\J$ ($q\ne 2$) and the global $\vartheta_2$, dubbed ``NH-SEA potentials,'' are the solution of the system of equations obtained by substituting Eq.~(\ref{eq:NHSEA_general}) into the conservation constraints, $\dot{c_q}|_\J=0$ for $q\ne 2$ and $\sum_{\J=1}^M\, \dot{c_2}|_\J =0$, such that
\begin{equation}\label{eq:multipliersNH1} 
	\Big((C_\ell)^\J_\rho\gamma_\J\Big|\hat G_\J^{-1}\Big|(M)^\J_\rho\gamma_\J\Big)=0\quad \forall \J\text{ and } \ell\ne 2\,,
\quad 
	\sum_{\J=1}^\M \Big((C_2)^\J_\rho\gamma_\J\Big|\hat G_\J^{-1}\Big|(M)^\J_\rho\gamma_\J\Big)=0\,.
\end{equation}

The same set of assumptions detailed in Appendices \ref{sec:seaqt} and \ref{sec:heseaqt} and Section \ref{sec:CSHE} are adopted with regard to (i) the local metrics $G_\J$ (Assumptions SEAQT5-7); (ii) the absence of interaction terms in the overall system's Hamiltonian and correlations between subsystems (Assumption CSHE1); (iii) the HE decomposition of the Hilbert space of each subsystem (Assumptions CSHE2-4); (iv) the CSHE assumption (CSHE5) on the state; (v) the minimal set of generators of the motion, i.e., $Q=2$, $C_1=I$ and $C_2=H$, and the corresponding renaming of the Lagrange multipliers, $\vartheta_1^\J=\Boltz\alpha_\J$ and $\vartheta_2=\Boltz\beta $. As a result,
\begin{equation}\label{eq:MassieuHESAENH} 	 
	(M)^\J_\rho =  M_\J = S_\J-\Boltz\alpha_\J I_\J -\Boltz\beta H_\J =\Boltz
	\SumKJ [  (\alphaKJ- \alpha_\J)    \PKJ +(\betaKJ- \beta) \HKJ  ]\, , 
\end{equation} 
and the system of equations that determines the Lagrange multipliers becomes
\begin{equation}\label{eq:multiplierssuperHeat3} 
	\SumKJ \frac{\pKJ}{\tauKJ}\Tr(\rhoKJtilde M_\J)=0\,,\quad \forall \J\,,
	\qquad  	\sum_{\J=1}^\M\SumKJ \frac{\pKJ}{\tauKJ}\Tr(\rhoKJtilde H_\J M_\J)=0\,,
\end{equation}
It may be rewritten as
\begin{align}\label{eq:SEA_constraintsHEHeat} 
	\SumK \frac{1}{\tauKJ} \big[  (\alphaKJ- \alpha_\J)   \pKJ +(\betaKJ- \beta) \langle \HKJ \rangle \big]&=0\quad \forall \J\,,\\
\sum_{\J=1}^\M\SumK \frac{1}{\tauKJ} \big[  (\alphaKJ- \alpha_\J)   \langle \HKJ \rangle +(\betaKJ- \beta) \langle \HKJ\HKJ \rangle \big]&=0\,,
\end{align}
and has the solution 
\begin{align}\label{eq:multipliers_singleHE1NH} 
	\Boltz\alpha_\J &=\langle s \rangle_w^\J - \Boltz\beta\,\langle \varepsilon  \rangle_w^\J \,,\quad \forall \J\,,\\
	\Boltz\beta &= \frac{\displaystyle \sum_{\J=1}^\M\frac{1}{\Tau_\J}\Big(B^\J_{SH}   - B^\J_{S} B^\J_{H}\Big)}{\displaystyle \sum_{\J=1}^\M\frac{1}{\Tau_\J}\Big(B^\J_{HH}-B^\J_{H} B^\J_{H}\Big)}= \frac{\displaystyle \sum_{\J=1}^\M\frac{1}{\Tau_\J}
		\langle \Delta_w s\Delta_w\varepsilon \rangle_w^\J}{\displaystyle \sum_{\J=1}^\M\frac{1}{\Tau_\J}\langle \Delta_w \varepsilon\Delta_w\varepsilon \rangle_w^\J}=\Boltz\frac{\displaystyle \sum_{\J=1}^\M v_\J\,\beta_\J^\text{eff}}{\displaystyle \sum_{\J=1}^\M v_\J}\,,\label{eq:multipliers_singleHE2NH}\\
\Boltz\beta_\J^\text{eff}&=\frac{B^\J_{SH}   - B^\J_{S} B^\J_{H}}{B^\J_{HH}-B^\J_{H} B^\J_{H}}\,,\qquad  v_\J=\frac{B^\J_{HH}-B^\J_{H} B^\J_{H}}{\Tau_\J}\,,\label{eq:betaeffJ}
\end{align}
where the $B^\J$'s and the weighted averages $\langle \cdot  \rangle_w^\J$ are defined as in Eqs.~\eqref{eq:multipliers_singleHEB1}--\eqref{eq:multipliers_singleHEB3}.

The rates of change of the HE constraint potentials  are still given by relaxation equations like \eqref{eq:alphaKdot} and \eqref{eq:betaKdot}, and the subsystems' energies and entropies by relations similar to \eqref{eq:Edot1} and \eqref{eq:Sdot1}. Thus,
\begin{equation}
	\tauKJ\ddt{\alphaKJ}= \alpha_\J-\alphaKJ  \,,\qquad \tauKJ\ddt{\betaKJ}= \beta-\betaKJ  \,,
\end{equation}
\begin{equation}
	\dv{\langle \HKJ \rangle}{t}=- \langle \HKJ \rangle\dv{\alphaKJ}{t} -  \langle \HKJ\HKJ \rangle \dv{\betaKJ}{t}\, ,
\end{equation}
\begin{equation}
	\frac{1}{\Boltz}	\dv{\langle \SKJ\rangle}{t}= \Big(\pKJ- \langle \SKJ \rangle/\Boltz\Big)\dv{\alphaKJ}{t}+  \Big( \langle \HKJ \rangle-\langle \SKJ\HKJ \rangle/\Boltz\Big)\dv{\betaKJ}{t}\, ,
\end{equation}
These equations indicate that---while the overall system relaxes (with possible overshooting during the process) towards the stable equilibrium state in which all the $\betaKJ$'s have converged to a common value equal to $\beta$---the various sectors exchange energy and entropy, both within each subsystem  and across subsystems. In Section~\ref{sec:NH3}, this model is detailed for the case of  a composite of three systems $\A$, $\B$, and $\J$, where $\A$ and $\B$ are assumed to be in locally stable equilibrium states, and could be heat baths if their heat capacities are very large.

The same concept outlined in this section has been implemented for systems with variable amounts of constituents. In addition to globally constraining the mean energy, the mean number of particles of each type are globally constrained so that the dissipative term in the NH-SEAQT equation of motion results in  effective exchanges of energy and entropy as well as particles between subsystems. As discussed in~\cite{SpakovskyReynoldsAscencio2026}, this approach extends the modeling of heat, diffusion, and heat-and-mass-diffusion interactions to the nonequilibrium domain in which subsystems are in local equilibrium or in HES's that are not necessarily  close  to mutual equilibrium.

\section{NH-SEAQT Model of Heat Interaction Between Two Systems in Local but not Mutual Equilibrium}\label{sec:NH2}

To illustrate the applications allowed by the NH-SEAQT framework just outlined in Section~\ref{sec:heseaqt_heat}, the simplest case of  a composite of only two subsystems $\A$ and $\B$ each with a single HE sector is considered (hence the subscript 1, used below for notational consistency).  Under these conditions, the general relations of Section~\ref{sec:heseaqt_heat} reduce to
\begin{equation}\alpha^\A_1=\ln Z^\A_1(\beta^\A_1)\,,\quad \dv{\alpha^\A_1}{t}=-\langle H_\A\rangle \dv{\beta^\A_1}{t}\,,\quad \alpha^\B_1=\ln Z^\B_1(\beta^\B_1)\,,\quad \dv{\alpha^\B_1}{t}=-\langle H_\B\rangle \dv{\beta^\B_1}{t}\,,
\end{equation}
\begin{equation}\label{eq:beta_heat_AB}
	\beta=  \frac{v_\A\beta_1^\A + v_\B\beta_1^\B}{v_\A + v_\B}\,,\qquad v_\A=\frac{B^\A_{HH}-B^\A_{H} B^\A_{H}}{\Tau_\A}\,, \quad v_\B=\frac{B^\B_{HH}-B^\B_{H} B^\B_{H}}{\Tau_\B}\,,
\end{equation}
\begin{equation}	
 \tau_\A\ddt{\beta_1^\A}= \beta-\beta_1^\A\,,\qquad \tau_\B\ddt{\beta_1^\B}= \beta-\beta_1^\B  \,,
\end{equation}
\begin{equation}
	\dv{\langle H_\A \rangle}{t}= (\beta_1^\A-\beta)v_\A=-\frac{v_\A v_\B}{v_\A+ v_\B} \big(\beta_1^\B-\beta_1^\A\big)=  -(\beta_1^\B-\beta)v_\B=-	\dv{\langle H_\B \rangle}{t}\, ,
\end{equation}
\begin{equation}
		\frac{1}{\Boltz}\dv{\langle S_\A \rangle}{t}= (\beta_1^\A-\beta)v_\A\beta_1^\A\, , \qquad	\frac{1}{\Boltz}\dv{\langle S_\B \rangle}{t}= (\beta_1^\B-\beta)v_\B\beta_1^\B\, ,
\end{equation}
\begin{equation}
	\frac{1}{\Boltz}\dv{\langle S \rangle}{t}=\frac{v_\A v_\B}{v_\A+ v_\B} \big(\beta_1^\B-\beta_1^\A\big)^2\,\quad\text{(clearly $\ge 0$)}\, .
\end{equation}
The overall entropy production is non-negative until $\beta_1^\A$ and $\beta_1^\B$ equalize. Energy flows from $\A$ to $\B$ when $\beta_1^\A<\beta_1^\B$. Using standard thermodynamic notation (see, e.g., \cite{Beretta2026, callen1985thermodynamics}), it is denoted as $\dot E^{\A\to\B}$ and is taken to be positive in the direction of the arrow. 

An effective temperature $T_Q^{\A\B}$---where the subscript $Q$ indicates a quantity associated with the heat interaction---can also be identified and an entropy flow related to the energy flow via the heat-interaction expression defined, i.e.,
\begin{equation} 
    \dot S^{\A\to\B}=\frac{ \dot E^{\A\to\B}}{T_Q^{\A\B}}\,,\qquad \text{ with } T_Q^{\A\B}=\frac{1}{\Boltz\beta} \text{ where } \beta=  \frac{v_\A\beta_1^\A + v_\B\beta_1^\B}{v_\A + v_\B}\,.
\end{equation}
Here $T_Q^{\A\B}=1/\Boltz\beta$ gives physical meaning to the SEA nonequilibrium potential $\beta$, namely, that it is the weighted average of the inverse temperatures of the two interacting systems, as defined in Eq.~\eqref{eq:beta_heat_AB}, where the weights $v_\A=c_\A\big/\Boltz\beta_1^\A\tau_\A$ and $v_\B=c_\B\big/\Boltz\beta_1^\B\tau_\B$ change in time and are related to the heat capacities and relaxation times of the respective systems. With this identification of the entropy flow, the local rates of entropy production within the two systems can be identified and the energy and entropy balance equations written as
\begin{equation}
	\dv{\langle H_\A \rangle}{t}=-\dot E^{\A\to\B}\,,\qquad 	\dv{\langle H_\B \rangle}{t}=\dot E^{\A\to\B}
\end{equation}
\begin{equation}\label{eq:fourierAB}
	  \dot E^{\A\to\B}=\frac{v_\A v_\B}{v_\A+ v_\B} \big(\beta_1^\B-\beta_1^\A\big)\,,
\end{equation}
\begin{equation}
	\dv{\langle S_\A \rangle}{t}=-\Boltz\beta\,\dot E^{\A\to\B}+\dot S^\A_\text{irr}\,,\qquad 	\dv{\langle S_\B \rangle}{t}=\Boltz\beta\,\dot E^{\A\to\B}+\dot S^\B_\text{irr}\,,
\end{equation}
\begin{equation}  \dot S^\A_\text{irr}=\Boltz\big(\beta_1^\A-\beta\big)^2v_\A \,\quad\text{(clearly $\ge 0$)}\,, \qquad  \dot S^\B_\text{irr}=\Boltz\big(\beta_1^\B-\beta\big)^2v_\B \,\quad\text{(clearly $\ge 0$)} \,.
\end{equation}
Notice that Eq.~\eqref{eq:fourierAB} is cast as a Fourier-law like linear-looking relation between the energy flow and the finite difference in inverse temperatures of the two systems. It is, however, a highly nonlinear relation since the proportionality coefficient  $v_\A v_\B/(v_\A+ v_\B) = 
c_\A c_\B /(\Boltz \beta_1^\A \tau_\A c_\A + \Boltz \beta_1^\B \tau_\B c_\B)  $ is a nonlinear function of the inverse temperatures.

In the near-equilibrium limit as $\beta_1^\A$ and $\beta_1^\B$ approach each other so that $\beta_1^\A\approx\beta\approx\beta_1^\B$; the model is consistent with the strict definition of a heat interaction at temperature $T_Q^{\A\B}$---as given in \cite{GB1991} (Section 12.3) and \cite{Beretta2026} (Section 40)---as well as with the linear Fourier-like law with coefficient $v_\A v_\B/(v_\A+ v_\B) \approx 
c_\A c_\B T_Q^{\A\B} /(\tau_\A c_\A +  \tau_\B c_\B)  $.

\section{NH-HE--SEAQT Model of SEA-Driven Energy and Entropy Exchange between  a System and Two Other Systems in Local but not Mutual Equilibrium}\label{sec:NH3}

As a further illustration, consider the case of  a composite of three subsystems $\A$, $\B$, and $\J$, where $\A$ and $\B$ are each assumed to have a single HE sector (hence the subscript 1) (and could represent heat baths if their heat capacities are very large), while system $\J$ is assumed to be in HES's with respect to a HESS decomposition. The additional assumptions are: (i) uncorrelated states (i.e., $\rho=\rho_\A\otimes\rho_\J\otimes\rho_\B$); (ii) no interaction Hamiltonians (i.e., $V_{\J{-}\A}=0$, $V_{\J{-}\B}=0$, and $V_{\A{-}\B}=0$); (iii) time-independent Hamiltonians for $\A$ and $\B$ (i.e., $\inlineddt{H_\A}=0$ and  $\inlineddt{H_\B}=0$, so that the only way the composite system can interact with other systems---such as a work element---is via the time dependence of control parameters in the Hamiltonian operator $H_\J$). 

The Hilbert space, overall Hamiltonian operator, and overall state operators of the composite system are
\begin{equation}
	\Hil_\text{tot} =\Hil_\A \otimes \Hil_\J\otimes \Hil_\B\,,\qquad \Hil_\J= \bigoplus_{\smallK=1}^{M} \Hil_\smallK\,,
\end{equation} 
\begin{equation}
	H=   H_\A  \otimes  I_\J \otimes  I_\B + I_\A  \otimes  H_\J \otimes  I_\B+I_\A  \otimes  I_\J \otimes  H_\B\, , \qquad H_\J=\SumK \HKJ
\end{equation}
\begin{equation}
	\rho=\rho_\A\otimes\rho_\J\otimes\rho_\B\,,\qquad \rho_\A=\gamma_\A\gamma_\A^\dagger\,,\qquad \rho_\J=\gamma_\J\gamma_\J^\dagger\,,\qquad \rho_\B=\gamma_\B\gamma_\B^\dagger
\end{equation}
\begin{equation}
	\rho_\A= \frac{\exp(-\beta_1^\A  H_\A  )}{Z_\A (\beta_1^\A )}\,,\quad  \rho_\J=\SumK 
	\frac{\PK\exp(-\alphaK \PK -\betaK  \HK  )\PK}{\ZK (\betaK )}\,,\quad \rho_\B=\frac{\exp(-\beta_1^\B  H_\B  )}{Z_\B (\beta_1^\B )} \, .
\end{equation}

The key assumption that distinguishes this SEA model from that discussed in Appendix~\ref{sec:seaqt} is that the equation of motion is obtained from a less constrained variational principle than (SEAQT3). Instead of a local maximization problem for each subsystem (see Eq.~\eqref{eq:variational}), a single global entropy production maximization problem for the overall system is assumed, subject to the following nine constraints: (i) normalization for each subsystem (three constraints); (ii) mean energy conservation for each interacting pair $\J$-$\A$ and $\J$-$\B$ (two constraints); (iii) a direction constraint for each separate dissipative contribution $\dot\gamma^{d}_\A$, $\dot\gamma^{d\A}_\J$, $\dot\gamma^{d\B}_\J$, $\dot\gamma^{d}_\B$ (four constraints) where 
\begin{equation}
	\dv{\rho_\A}{t}~=~ \dot\gamma^{d}_\A\gamma^\dagger_\A+\gamma_\A \dot\gamma^{d\dagger}_\A \,,\qquad \dv{\rho_\B}{t}~=~ \dot\gamma^{d}_\B\gamma^\dagger_\B+\gamma_\B \dot\gamma^{d\dagger}_\B \,, 
\end{equation}
\begin{equation}
	\dv{\rho_\J}{t}~=~ \big(\dot\gamma^{d\A}_\J+\dot\gamma^{d\B}_\J\big)\gamma^\dagger_\J+\gamma_\J \big(\dot\gamma^{d\A\dagger}_\J + \dot\gamma^{d\B\dagger}_\J\big)\,
\end{equation}
and therefore $\dot\gamma^{d}_\A$, $\dot\gamma^{d\A}_\J$, $\dot\gamma^{d\B}_\J$, $\dot\gamma^{d}_\B$ are given by the solution of the following maximization problem (in terms of the nine Lagrange multipliers $\Boltz\alpha_\A$, $\Boltz\alpha_\B$,  $\Boltz\alpha_\J$, $\Boltz\beta_\J^\A$, $\Boltz\beta_\J^\B$, $2\Boltz\tau_\A$, $2\Boltz\tau_\B$, $\Boltz\tau^\A_\J/2$, $\Boltz\tau^\B_\J/2$):
\begin{align}\label{eq:variationalHeat} 
    &	\max_{\dot\gamma^{d}_\A,\dot\gamma^{d\A}_\J,\dot\gamma^{d\B}_\J,\dot\gamma^{d}_\B}\ \Upsilon= \Big(2\gamma_\A S_\A\Big|\dot\gamma^{d}_\A\Big)+\Big(2\gamma_\J S_\J\Big|(\dot\gamma^{d\A}_\J+\dot\gamma^{d\B}_\J)\Big)+\Big(2\gamma_\B S_\B\Big|\dot\gamma^{d}_\B\Big) \nonumber\\ &-\Boltz\alpha_\A\Big(2\gamma_\A\Big|\dot\gamma^{d}_\A\Big)- \Boltz\alpha_\J\Big(2\gamma_\J\Big|(\dot\gamma^{d\A}_\J+\dot\gamma^{d\B}_\J)\Big) -\Boltz\alpha_\B\Big(2\gamma_\B\Big|\dot\gamma^{d}_\B\Big)\nonumber\\
	&- \Boltz\beta_\J^\A\,\Big[\Big(2\gamma_\A H_\A\Big|\dot\gamma^{d}_\A\Big)+\Big(2\gamma_\J H_\J\Big|\dot\gamma^{d\A}_\J\Big)\Big]- \Boltz\beta_\J^\B\,\Big[\Big(2\gamma_\J H_\J\Big|\dot\gamma^{d\B}_\J\Big)+\Big(2\gamma_\B H_\B\Big|\dot\gamma^{d}_\B\Big)\Big]\nonumber\\
	&-2\Boltz\tau_\A \Big(\dot\gamma^d_\A\Big|\dot\gamma^d_\A\Big) -\frac{\Boltz\tau^\A_\J}{2} \Big(\dot\gamma^{d\A}_\J\Big|\,\hat G_\J\,\Big|\dot\gamma^{d\A}_\J\Big)-\frac{\Boltz\tau^\B_\J}{2} \Big(\dot\gamma^{d\B}_\J\Big|\,\hat G_\J\,\Big|\dot\gamma^{d\B}_\J\Big)-2\Boltz\tau_\B \Big(\dot\gamma^d_\B\Big|\dot\gamma^d_\B\Big)\,,
\end{align}
The last four constraints correspond to the conditions necessary for maximizing with respect to  local directions only. They are computed for systems $\A$ and $\B$ with respect to a uniform Fisher-Rao metric and for the two contributions $\J$-$\A$ and $\J$-$\B$  with respect to a metric compatible with assumptions SEAQT5 of Appendix \ref{sec:seaqt} and HE--SEAQT6 and HE--SEAQT7 of Appendix~\ref{sec:heseaqt}.

A distinctive feature of this maximization problem is the double energy conservation constraint: one ensuring that the $\dot\gamma^{d\A}_\J$ contribution conserves the overall $\A$+$\J$ mean energy  and the other that the $\dot\gamma^{d\B}_\J$ contribution conserves the overall $\J$+$\B$ mean energy. This less restrictive hybrid assumption is crucial because it results in non-Hamiltonian energy exchanges between $\A$ and $\J$ and between $\J$ and $\B$ but not directly between $\A$ and $\B$.

Taking the variational derivatives of $\Upsilon$ with respect to $|\dot\gamma^{d}_\A)$, $|\dot\gamma^{d\A}_\J)$,  $|\dot\gamma^{d\B}_\J)$, and $|\dot\gamma^{d}_\B)$ and setting them equal to zero yields, in terms of the local  nonequilibrium Massieu operators,
\begin{align}
	\frac{\delta\Upsilon}{|\delta\dot\gamma^{d}_\A)}&= \Big|2M_\A\gamma_\A\Big) -4\Boltz\tau_\A\, \Big|\dot\gamma^d_\A\Big)=0\,,& 	M_\A&=S_\A-\Boltz\alpha_\A I_\A -\Boltz\beta_\J^\A\, H_\A \,,\\
	\frac{\delta\Upsilon}{|\delta\dot\gamma^{d\A}_\J)}&= \Big|2 M^\A_\J\gamma_\J\Big) -\Boltz\tau^\A_\J\, \hat G_\J\,\Big|\dot\gamma^{d\A}_\J\Big)=0\,,& M^\A_\J&=S_\J-\Boltz\alpha_\J I_\J -\Boltz\beta_\J^\A\, H_\J\,,\\
	\frac{\delta\Upsilon}{|\delta\dot\gamma^{d\B}_\J)}&= \Big|2 M^\B_\J\gamma_\J\Big) -\Boltz\tau^\B_\J\, \hat G_\J\,\Big|\dot\gamma^{d\B}_\J\Big)=0\,,& M^\B_\J&=S_\J-\Boltz\alpha_\J I_\J -\Boltz\beta_\J^\B\, H_\J\,,\\
	\frac{\delta\Upsilon}{|\delta\dot\gamma^{d}_\B)}&= \Big|2M_\B\gamma_\B\Big) -4\Boltz\tau_\B\, \Big|\dot\gamma^d_\B\Big)=0\,,&	M_\B&=S_\B-\Boltz\alpha_\B I_\B -\Boltz\beta_\J^\B\, H_\B \,,
\end{align}
and, therefore,
\begin{align} 
	\Big|\dot\gamma^d_\A\Big)&=\frac{1}{2\Boltz\tau_\A}\Big|M_\A\gamma_\A\Big)\,,&	\Big|\dot\gamma^{d\A}_\J\Big)&=\frac{1}{\Boltz\tau^\A_\J}\hat G_\J^{-1}	\Big|2M^\A_\J\gamma_\J\Big)\,,\label{eq:SEA_general_3body_gammaA}\\
    \Big|\dot\gamma^d_\B\Big)&=\frac{1}{2\Boltz\tau_\B}\Big|M_\B\gamma_\B\Big)\,,& \Big|\dot\gamma^{d\B}_\J\Big)&=\frac{1}{\Boltz\tau^\B_\J}\hat G_\J^{-1}	\Big|2M^\B_\J\gamma_\J\Big)\,,\label{eq:SEA_general_3body_gammaB}
\end{align}
The Lagrange multipliers $\alpha_\A$, $\alpha_\J$, $\alpha_\B$, $\beta_\J^\A$, $\beta_\J^\B$ are found from the solution of the system of equations obtained by substituting Eqs.~\eqref{eq:SEA_general_3body_gammaA} and \eqref{eq:SEA_general_3body_gammaB} into the conservation constraints such that 
\begin{equation} 
	\Big(\gamma_\A\Big|M_\A\gamma_\A\Big)=0\,,\quad \Big(\gamma_\J\Big|\hat G_\J^{-1}\Big|\big(M^\A_\J/\tau^\A_\J +M^\B_\J/\tau^\B_\J \big)\gamma_\J\Big)=0\,,\quad \Big(\gamma_\B\Big|M_\B\gamma_\B\Big)=0\,,
\end{equation}
\begin{equation}
	\frac{1}{\tau_\A}\Big(\gamma_\A H_\A\Big|M_\A\gamma_\A\Big)+\frac{4}{\tau^\A_\J}\Big(\gamma_\J H_\J\Big|\hat G_\J^{-1}\Big|M^\A_\J\gamma_\J\Big)=0\,. 
\end{equation}
\begin{equation}
	\frac{4}{\tau^\B_\J}\Big(\gamma_\J H_\J\Big|\hat G_\J^{-1}\Big|M^\B_\J\gamma_\J\Big)+\frac{1}{\tau_\B}\Big(\gamma_\B H_\B\Big|M_\B\gamma_\B\Big)=0\,. 
\end{equation}

Under the stated HE assumptions (SEAQT5, HE--SEAQT6, HE--SEAQT7), the above equations reduce to the following for the local density operators and are similar to Eq.~\eqref{eq:SEA_generalHE} except for the double $\beta$'s, i.e., the two SEA  potentials $\beta^\A_\J$ and $\beta^\B_\J$ instead of a single one:
\begin{align}
	\dv{\rho_\A}{t}&= \frac{1}{\tau_\A} \big[  \big(\alpha_1^\A- \alpha_\A\big)    \rho_\A +(\beta_1^\A- \beta^\A_\J) H_\A \rho_\A \big]\,,\label{eq:rhoAdot3}\\
	\dv{\rho_\J}{t}&=\frac{\Tau_\J}{\tau^\A_\J}\SumK \frac{\pK}{\tauK} \big[  (\alphaK- \alpha_\J)    \rhoKtilde +(\betaK- \beta^\A_\J) \HK \rhoKtilde \big]\nonumber\\&+\frac{\Tau_\J}{\tau^\B_\J}\SumK \frac{\pK}{\tauK} \big[  (\alphaK- \alpha_\J)    \rhoKtilde +(\betaK- \beta^\B_\J) \HK \rhoKtilde \big]\,,\label{eq:rhoJdot3}\\
	\dv{\rho_\B}{t}&= \frac{1}{\tau_\B} \big[  \big(\alpha_1^\B- \alpha_\B\big)    \rho_\B +(\beta_1^\B- \beta^\B_\J) H_\B \rho_\B \big]\,.\label{eq:rhoBdot3}
\end{align}
Using the same procedure as in the derivation of Eqs.~\eqref{eq:pikdotHES} and \eqref{eq:sikdotHES} from Eq.~\eqref{eq:SEA_singleHE}, Eq.~\eqref{eq:rhoJdot3} together with $\piK\giK=\Tr(\rho\PiK)$ and $\siK=-\Boltz\ln\piK$  yields the following relations:
\begin{equation}\label{eq:pikdotHES3}
    \tauK	\ddt{\piK }= (\omega_\J^\A+\omega_\J^\B)\big[(\alphaK -\alpha_\J) \piK +(\betaK -\beta_\J^{\A\B})\,\eiK \piK\big] \,,
\end{equation}
\begin{equation}\label{eq:sikdotHES3}
    \tauK	\ddt{\siK }= (\omega_\J^\A+\omega_\J^\B)\big[(\alphaK -\alpha_\J)  +(\betaK -\beta_\J^{\A\B})\,\eiK \big]    \,,
\end{equation}
where
\begin{equation}
	\omega_\J^\A = \frac{\Tau_\J}{\tau^\A_\J} \,, \quad \omega_\J^\B = \frac{\Tau_\J}{\tau^\B_\J} \,,\qquad \beta_\J^{\A\B}=\frac{\omega_\J^\A \beta^\A_\J + \omega_\J^\B \beta^\B_\J}{\omega_\J^\A + \omega_\J^\B}\,,
\end{equation}
so that the equivalent expressions of Eqs.~\eqref{eq:alphaKdot} and \eqref{eq:betaKdot} become for the present case
\begin{equation}\label{eq:alphabetadot3}
	\frac{1}{\omega_\J^\A + \omega_\J^\B}\ddt{\alphaK}= \frac{\alpha_\J-\alphaK}{\tauK}  \,,\qquad \frac{1}{\omega_\J^\A + \omega_\J^\B}\ddt{\betaK}= \frac{\beta_\J^{\A\B}-\betaK}{\tauK}  \,.
\end{equation}

The system of equations that determines the Lagrange multipliers $\alpha_\A$, $\alpha_\J$, $\alpha_\B$, $\beta_\J^\A$, $\beta_\J^\B$ is then written as
\begin{equation}
	\Tr(\rho_\A M_\A)=0\,,\quad 	\SumK \frac{\pK}{\tauK}\Tr\Big(\rhoKtilde \big(M^\A_\J/\tau^\A_\J +M^\B_\J/\tau^\B_\J \big)\Big)=0\,,\quad 	\Tr(\rho_\B M_\B)=0\,,
\end{equation}
\begin{equation}
	\frac{1}{\tau_\A}\Tr(\rho_\A H_\A M_\A)+\frac{\Tau_\J}{\tau^\A_\J}\SumK \frac{\pK}{\tauK}\Tr(\rhoKtilde H_\J M^\A_\J)=0\, 
\end{equation}
\begin{equation}
	\frac{\Tau_\J}{\tau^\B_\J}\SumK \frac{\pK}{\tauK}\Tr(\rhoKtilde H_\J M^\B_\J)+\frac{1}{\tau_\B}\Tr(\rho_\B H_\B M_\B)=0\, 
\end{equation}
and, using Eqs.~\eqref{eq:alphabetadot3}, may be rewritten as
\begin{equation}
	\big(\alpha_1^\A- \alpha_\A\big)    +(\beta_1^\A- \beta^\A_\J) \langle H_\A \rangle=0\,,\qquad
	\big(\alpha_1^\B- \alpha_\B\big)    +(\beta_1^\B- \beta^\B_\J) \langle H_\B \rangle=0\,,
\end{equation}
\begin{equation}\label{eq:sys1}
	\SumK\Big[ \frac{\alphaK- \alpha_\J}{\tauK}   \pK +  \frac{\betaK- \beta^{\A\B}_\J}{\tauK} \langle \HK \rangle \Big] =0\,,
\end{equation}
\begin{equation}\label{eq:sys2}
\frac{\alpha_1^\A- \alpha_\A}{\tau_\A} \langle H_\A  \rangle +	\frac{\beta_1^\A- \beta^\A_\J}{\tau_\A} \langle H_\A H_\A \rangle + \omega_\J^\A \SumK\Big[ \frac{\alphaK- \alpha_\J}{\tauK}   \langle \HK \rangle +  \frac{\betaK- \beta^\A_\J}{\tauK} \langle \HK\HK \rangle\Big] =0\,,
\end{equation}
\begin{equation}\label{eq:sys3}
		\frac{\alpha_1^\B- \alpha_\B}{\tau_\B} \langle H_\B  \rangle +	\frac{\beta_1^\B- \beta^\B_\J}{\tau_\B} \langle H_\B H_\B \rangle +	\omega_\J^\B \SumK\Big[ \frac{\alphaK- \alpha_\J}{\tauK}   \langle \HK \rangle + 	\frac{\betaK- \beta^\B_\J}{\tauK} \langle \HK \HK \rangle\Big]=0\,.
\end{equation}
Recalling the definitions of $\beta_\J^\text{eff}$ and $v_\J$ (Eq.~\eqref{eq:betaeffJ}),
\begin{equation}
 v_\J=\frac{B^\J_{HH}-B^\J_{H} B^\J_{H}}{\Tau_\J}\, \,,\qquad
\beta_\J^\text{eff}=\frac{1}{\Boltz} \frac{B^\J_{SH}   - B^\J_{S} B^\J_{H}}{\Tau_\J\, v_\J} \,,
\end{equation}
and defining 
 \begin{equation}
	v_\A=\frac{ \langle H_A H_A \rangle- \langle H_A  \rangle \langle H_A  \rangle}{\tau_\A}\,,\qquad 	v_\B=\frac{ \langle H_B H_B \rangle- \langle H_B  \rangle \langle H_B  \rangle}{\tau_\B}\,,
\end{equation}
 the solution to the system of Eqs. \eqref{eq:sys1}, \eqref{eq:sys2}, and \eqref{eq:sys3}  for the multipliers $\alpha_\J, \beta^\A_\J, \beta^\B_\J$ is given by
\begin{equation}
	\alpha_\J = B_S^\J - \beta_\J^{\A\B} B_H^\J\,, \qquad
		\beta^\A_\J = \frac{v_\A\beta_1^\A+v_\J\beta_\J^\text{eff}}{v_\A+v_\J}\,, \qquad \beta^\B_\J = \frac{v_\J\beta_\J^\text{eff}+v_\B\beta_1^\B}{v_\J+v_\B}\,.
\end{equation}

The rate of change of the energy of $\J$ is now expressed as
\begin{equation}
	\dv{\langle H_\J \rangle}{t} = \frac{v_\A v_\J}{v_\A + v_\J} (\beta_\J^{\text{eff}} - \beta_1^\A) + \frac{v_\B v_\J}{v_\B + v_\J} (\beta_\J^{\text{eff}} - \beta_1^\B)\,,
\end{equation}
 and the energy balance equations for the three subsystems  (recall that the notation adopted is \cite{Beretta2026}, $\dot E^{\A\to\J}=-\dot E^{\J\to\A}$ and $\dot E^{\B\to\J}=-\dot E^{\J\to\B}$) are
\begin{equation}
	\dv{\langle H_\A \rangle}{t}= -\dot E^{\A\to\J}\, ,\qquad
	\dv{\langle H_\J \rangle}{t}=\dot E^{\A\to\J}- \dot E^{\J\to\B}\, ,\qquad
	\dv{\langle H_\B \rangle}{t}= \dot E^{\J\to\B}\, ,
\end{equation}
\begin{equation}
	\dot E^{\A\to\J}= (\beta_\J^\A-\beta_1^\A)v_\A=(\beta_\J^\text{eff}-\beta_\J^\A)v_\J=\frac{v_\A v_\J}{v_\A+ v_\J} \big(\beta_\J^\text{eff}-\beta_1^\A\big)\, ,
\end{equation}
\begin{equation}
	\dot E^{\J\to\B}= (\beta_1^\B-\beta_\J^\B)v_\B=(\beta_\J^\B-\beta_\J^\text{eff})v_\J=-\frac{v_\B v_\J}{v_\B+ v_\J} \big(\beta_\J^\text{eff}-\beta_1^\B\big)\, .
\end{equation}
The rates of change of the entropy of $\A$, $\B$, and $\J$ are given by
\begin{equation}
    \frac{1}{\Boltz}\dv{\langle S_\A \rangle}{t}=\beta_1^\A	\,\dv{\langle H_\A \rangle}{t} \, ,\qquad
    \frac{1}{\Boltz}\dv{\langle S_\B \rangle}{t}=\beta_1^\B\,	\dv{\langle H_\B \rangle}{t} \, ,
\end{equation}
\begin{equation}\label{eq:SJDOT3}
    \frac{1}{\Boltz}\dv{\langle S_\J \rangle}{t} = \Big[\frac{v_\A v_\J}{v_\A + v_\J} (\beta_\J^{\text{eff}} - \beta_1^\A) + \frac{v_\B v_\J}{v_\B + v_\J} (\beta_\J^{\text{eff}} - \beta_1^\B)\Big]\beta_\J^{\text{eff}}= \beta_\J^{\text{eff}}\,	\dv{\langle H_\J \rangle}{t}  \,,
\end{equation}
which identifies  $\Boltz\beta_\J^{\text{eff}}$  as the effective inverse temperature of the HE system $\J$,  relating its energy and entropy changes through a stable-equilibrium Gibbs-like relation.

Two other effective temperatures, $T_Q^{\A\J}=1/\Boltz\beta_\J^\A $ and $T_Q^{\J\B}=1/\Boltz\beta_\J^\B $, define the ratio of energy to entropy flows between $\J$ and $\A$, and between $\J$ and $\B$, respectively, via the typical heat-interaction expressions expressed as
\begin{equation}
    \dot S^{\A\to\J}=\Boltz\beta^\A_\J \dot E^{\A\to\J}=\frac{ \dot E^{\A\to\J}}{T_Q^{\A\J}}\,,\qquad 
    \dot S^{\J\to\B}=\Boltz\beta^\B_\J \dot E^{\J\to\B}=\frac{ \dot E^{\J\to\B}}{T_Q^{\J\B}}\,,
\end{equation}
As a result, the following consistent  entropy balance equations for the three subsystems are written as
\begin{equation}
    \dv{\langle S_\A \rangle}{t}= -\dot S^{\A\to\J} + \dot S^\A_\text{irr}\,,\qquad 
    \dv{\langle S_\B \rangle}{t}= \dot S^{\J\to\B} + \dot S^\B_\text{irr}\,,
\end{equation}
\begin{equation}
    \dv{\langle S_\J \rangle}{t}= \dot S^{\A\to\J} - \dot S^{\J\to\B} + \dot S^\J_\text{irr}\,,
\end{equation}
where the expressions for the  entropy generation rates in the three systems, clearly $\ge 0$, are
\begin{equation}
    \frac{1}{\Boltz}\dot S^\A_\text{irr} = (\beta_1^\A-\beta_\J^\A)^2 v_\A=  \frac{v_\A v_\J^2 }{(v_\A + v_\J)^2} (\beta_\J^{\text{eff}} - \beta_1^\A)^2\,,
\end{equation}
\begin{equation}
    \frac{1}{\Boltz}\dot S^\B_\text{irr} =(\beta_1^\B-\beta_\J^\B)^2 v_\B=  \frac{v_\B v_\J^2 }{(v_\B + v_\J)^2} (\beta_\J^{\text{eff}} - \beta_1^\B)^2,
\end{equation}
\begin{equation}
    \frac{1}{\Boltz}\dot S^\J_\text{irr} =  \frac{v_\A^2 v_\J }{(v_\A + v_\J)^2} (\beta_\J^{\text{eff}} - \beta_1^\A)^2 + \frac{v_\B^2 v_\J}{(v_\B + v_\J)^2} (\beta_\J^{\text{eff}} - \beta_1^\B)^2 \,.
\end{equation}

These relations, for example, show that a steady state for $\J$, defined by the condition that  $\dhv{\langle H_\J \rangle}{t}=0$ and $\dhv{\langle S_\J \rangle}{t}=0$,  requires  that $\beta_\J^\text{eff}$ obey the following weighted sum of the inverse temperatures of $\A$ and $\B$: 
\begin{equation}
	\beta_{\J}^\text{eff}|_\text{s.s.}   = \frac{(v_\B + v_\J)v_\A\beta_1^\A + (v_\A + v_\J)v_\B\beta_1^B}{(v_\B + v_\J)v_\A + (v_\A + v_\J)v_\B} \left\{ 
	\begin{array}{l}
	\displaystyle	\xrightarrow{\quad v_\J\gg v_\A, v_\B\quad } \frac{v_\A\beta_1^\A + v_\B\beta_1^B}{v_\A +v_\B} \\[2ex]
	\displaystyle	\xrightarrow{\quad v_\J\ll v_\A, v_\B \quad } \frac{\beta_1^\A +\beta_1^B}{2}
	\end{array} \right. \, ,
\end{equation}
The two limiting cases---when $\J$ relaxes much faster than $\A$ and $\B$ and when $\J$ relaxes much more slowly---are highlighted by the limit expressions in that equation. 

Another important special case arises when $v_\A \gg v_\J$ and $v_\B \gg v_\J$. In this limit, $\beta_\J^\A\approx \beta_1^\A$, $\beta_\J^\B\approx \beta_1^\B$, $\dot S^\A_\text{irr}\approx 0$, $\dot S^\B_\text{irr}\approx 0$, and systems $\A$ and $\B$ model the behavior of heat baths. Furthermore, by adjusting the time dependence of the parameters $\omega_\J^\A$ and $\omega_\J^\B$ and the HE Hamiltonians $\HK$, the NH-HE--SEAQT equations \eqref{eq:rhoAdot3}–\eqref{eq:rhoBdot3} can model a quantum thermal machine $\J$ coupled to two reservoirs within a fully thermodynamically consistent framework (see, e.g., \cite{Li2016c,Militello2018}).

\section{Conclusions}\label{sec:conclusions}

In this work, a rigorous mathematical foundation for the HES concept within the framework of SEAQT is established. Using a general Hilbert space decomposition, a precise operator-level definition of HES's is provided, and the reduced dynamical equations for their associated intensive parameters derived. A central result of this work is the proof of the invariant-manifold property, which demonstrates that the SEAQT equation of motion preserves the $M$-th-order HE structure. This justifies the use of HE variables as a consistent reduced-order representation of full quantum dynamics, ensuring that an initial ``mixture of canonicals'' remains within its own family during evolution.

The HE--SEAQT framework is also extended to model composite systems via a NH--SEAQT approach.  Unlike standard models where energy exchange is restricted to interaction Hamiltonians, the NH--SEAQT approach uses an `entropic coupling'---a direct dissipative driving of energy exchange between subsystems via the less constrained SEA dissipative term. This allows for a thermodynamically consistent description of heat, diffusion, and heat-and-mass-diffusion interactions between subsystems even in the far-from-equilibrium domain. When mean energy and particle numbers are constrained globally, the NH--SEAQT equation effectively captures the exchange of energy, entropy, and constituents between subsystems that are not necessarily close to mutual equilibrium.

Finally, the theoretical positioning of the HE--SEAQT model is clarified by establishing its formal consistency with the rate-controlled constrained equilibrium (RCCE) method. This connection identifies the evolving HE parameters as physical constraint potentials, unifying the SEAQT dissipative structure with maximum-entropy principles.  These links  validate the HE--SEAQT approach as a robust, computationally efficient framework for reduced-order modeling of complex, far-from-equilibrium phenomena. The mathematical consistency demonstrated here supports its ongoing application to diverse problems in quantum transport, chemical kinetics, and microstructural evolution, providing a bridge between fundamental quantum dissipation and macroscopic nonequilibrium thermodynamics.



\appendix
\setcounter{section}{0}
\setcounter{equation}{0}

\renewcommand{\thesection}{\Alph{section}}

\renewcommand{\theequation}{\Alph{section}\arabic{equation}}

\makeatletter
\let\oldsection\section
\renewcommand{\section}[1]{\oldsection{\hspace{-0.5em}.\ #1}}
\makeatother

\section{SEAQT Equations of Motion}\label{sec:seaqt}

In this appendix, the tenets of the original SEAQT formalism are reviewed. A detailed discussion  can be found in the recent article~\cite{RayBeretta2025}, which also discusses the foundational issues of ``nosignaling'' and ``strong second-law compatibility'' that are necessary requirements of any nonlinear, nonlocal, nonequilibrium model of quantum dynamics. 

An essential ingredient of a nonlinear, dissipative quantum evolution equation for a general composite system  requires declaring the system's structure-dependent expressions so that the separate contribution of each subsystem to the dissipative term in the equation of motion for the overall state represented  by the density operator $\rho$ on the system's Hilbert space $\Hil=\bigotimes_{\J=1}^{\M}\Hil_\J$ can be determined. The internal structure of the system determines which of the $\M$ subsystems are to be protected from nonphysical effects such as signaling,  the exchange of energy, or the build-up of correlations between noninteracting subsystems. Using the notation introduced in~\cite{RayBeretta2025} for the dissipative term---which supplements the usual nondissipative Hamiltonian term---the following nosignaling structure is assumed: 

\begin{enumerate}[label=\textbf{(SEAQT\arabic*):}, labelsep=4pt, leftmargin=*]\item The SEAQT equation of motion for a general composite of quantum subsystems is given by
	\begin{equation}\label{eq:SEA_Composite_simp}
		\dv{\rho}{t}~=~-\frac{\imi}{\hbar}\comm{\mathit{H}}{\rho}-\sum_{\J=1}^\M\acomm*{\mathcal{D}^\J_\rho}{\rho_\J}\otimes\rho_\Jbar\,,
	\end{equation}
	where the $\J$-th subsystem's dissipation operator $\mathcal{D}^\J_\rho$ (on $\Hil_\J$) may be a nonlinear function of the local observables of $\J$, \xout{of} the reduced state $\rho_\J=\Tr_\Jbar(\rho)$, and \xout{of} the local perception operators (LPOs) of overall observables---where LPOs are operators on $\Hil_\J$ defined  as  $(X)^\J_\rho ~=~\Tr_\Jbar[(I_\J{\otimes}\rho_\Jbar)X]$
	with $\rho_\Jbar=\Tr_\J(\rho)$. Partial tracing Eq.~\eqref{eq:SEA_Composite_simp} over $\Hil_\Jbar$ yields 
    \begin{equation}\label{eq:SEA_local}
		\dv{\rho_\J}{t}~=~-\frac{\imi}{\hbar}\comm*{H_\J}{\rho_\J} -\frac{\imi}{\hbar}\Tr_\Jbar(\comm*{V}{\rho}) -\acomm*{\DrhoJ}{\rho_\J}\,,
	\end{equation}
	where $V$ is the interaction Hamiltonian. Note that the second term on the RHS can be expressed for weak interactions and under well-known assumptions in GKSL form.
	
	\item For the dissipative term to preserve $\Tr(\rho)$, operators $\acomm*{\DrhoJ}{\rho_\J}$ must be traceless. To preserve $\Tr(\rho H)$ (and possibly other conserved properties or charges, $\Tr(\rho C_q)$), operators  $\acomm*{\DrhoJ}{\rho_\J}(H)^\J_\rho$ (and $\acomm*{\DrhoJ}{\rho_\J}(C_q)^\J_\rho$, with $q=1\cdots Q$) must also be traceless. The rate of change of the overall system's entropy $\langle S\rangle =-\Boltz\Tr[\rho\ln\rho]$ is 
	\begin{equation}\label{eq:SEA_entropy_prod}
		\dv{\langle S\rangle}{t}= -\sum_{\J=1}^\M\Tr[\acomm*{\DrhoJ}{\rho_\J}(S)^\J_\rho]\,.
	\end{equation}
\end{enumerate}

As proven in \cite{RayBeretta2025}, for all possible choices of $\DrhoJ$, Eq.~\eqref{eq:SEA_Composite_simp} defines a broad class of nosignaling nonlinear evolution equations that are not restricted by the often assumed condition that ${\rm d}\rho_\J/{\rm d}t$ be a function of $\rho_\J$ only, which is sufficient but not necessary to prevent signaling.

The SEA assumption---in the spirit of the fourth law of thermodynamics~\cite{Beretta2020}---can be cast in terms of a variational principle which selects the dissipation operators $\DrhoJ$. As a first step, to trivially preserve the non-negativity and self-adjointness of $\rho$ during its time evolution, the generalized square root of $\rho_\J$, $\gamma_\J(t)=\sqrt{\rho_\J(t)}\,U$, is defined where $U$ is an arbitrary unitary operator such that
\begin{equation}\label{eq:gammaDef}
	\rho_\J=\gamma_\J\gamma_\J^\dagger. 
\end{equation}
The dissipative term in Eq.~(\ref{eq:SEA_Composite_simp}) is 
rewritten as
\begin{equation}\label{eq:SEA_Composite_gamma}
	-\acomm*{\DrhoJ}{\rho_\J}=\dot\gamma^d_\J\gamma_\J^\dagger+\gamma_\J\dot\gamma^{d\dagger}_\J\qquad \text{ with } \dot\gamma^d_\J=-\DrhoJ\gamma_\J \,.
\end{equation}

Next, on the set $\mathcal{L}(\Hil_\J)$ of linear operators on $\Hil_\J$, the  real inner product $( \cdot | \cdot )$ is defined as
\begin{equation}\label{eq:innerProduct}
	( X | Y)=\Tr (X^{\dagger} Y + Y^{\dagger} X )/2\,,
\end{equation}
so that the unit-trace condition for $\rho_\J$ is rewritten as $( \gamma_\J | \gamma_\J)=1$, implying that the $ \gamma_\J$'s lie on the unit sphere in $\mathcal{L}(\Hil_\J)$---along with their time-evolved trajectories, $\gamma_\J(t)$.
Along these trajectories, the distance traveled between $t$ and $t{\,+\,}\dd t$ can be expressed as
\begin{equation}\label{eq:distance}
	\dd \ell_\J~=~\sqrt{(\dot\gamma_\J|\,\hat G_\J(\gamma_\J)\,|\dot\gamma_\J)}\, \dd  t\,,
\end{equation}
where  $\hat G_\J(\gamma_\J)$ is some real, dimensionless, symmetric, and positive-definite operator on  $\mathcal{L}(\Hil_\J)$ (a superoperator on $\Hil$, possibly a nonlinear function of $\rho_\J$) that plays the role of a local metric tensor field characterizing the system's internal perception of  distance between nonequilibrium states.

The rates of change of the overall system entropy, $\langle S\rangle$ (Eq.~\eqref{eq:SEA_entropy_prod}), and of the overall system mean values of the $Q$ linear charges, $\langle C_q\rangle=\Tr(\rho C_q)$, can be written as
\begin{alignat}{2}
	\dv{\langle S\rangle}{t}&= \sum_{\J=1}^\M\dot{s}|_\J &\qquad \dot{s}|_\J&=\Big(2(S)^\J_\rho\gamma_\J\Big|\dot\gamma^{d}_\J\Big)\,, \label{eq:SEA_entropy_prod_gamma}\\
	\dv{\langle C_q\rangle}{t}&= \sum_{\J=1}^\M\dot{c_q}|_\J&\qquad \dot{c_q}|_\J&=\Big(2(C_q)^\J_\rho\gamma_\J\Big|\dot\gamma^{d}_\J\Big)\,, \label{eq:constants_gamma}
\end{alignat}
where both equations exhibit additive contributions from the subsystems.

Finally, the variational principle that leads to expressions for the $\dot\gamma^{d}_\J$ and the $\mathcal{D}^\J_\rho$  
that define the composite-system version of the SEA equation of motion is stated as follows: 

\begin{enumerate}[label=\textbf{(SEAQT\arabic*):}, labelsep=4pt,start=3, leftmargin=*]\item
	The  dissipative part of the evolution in Eq.~\eqref{eq:SEA_Composite_simp} ensures that, with respect to a local dissipative metric $\hat G_\J$, the direction of the local trajectory $\gamma_\J(t)$ maximizes the local contribution of  $\dot{s}|_\J$ to the overall system's entropy production rate; while the constraints $\dot{c_q}|_\J=0$  guarantee that the dissipative part of the dynamics does not contribute to the rates of change of the locally perceived  global charges $C_q$ (so that they emerge as constants of the motion if they also commute with the Hamiltonian, i.e., $\comm{H}{C_q}=0$).  \end{enumerate}

Introducing the Lagrange multipliers $\vartheta_q^\J$ and $\tau_\J$ for the constraints, the SEAQT $\dot\gamma^d_\J$'s are found by solving the maximization problem
\begin{equation}\label{eq:variational} 
	\max_{\dot\gamma^{d}_\J}\ \Upsilon_\J= \dot{s}|_\J - \sum_{q=1}^Q\vartheta_q^\J\, \dot{c_q}|_\J -\frac{\Boltz\tau_\J}{2} \Big(\dot\gamma^d_\J\Big|\,\hat G_\J\,\Big|\dot\gamma^d_\J\Big)\,,
\end{equation}
where the last constraint corresponds to the condition $(\dd \ell^d_\J/\dd  t)^2= \text{constant}$, necessary for maximizing with respect to direction only (see~\cite{Beretta2014,Gheorghiu2001a,Gheorghiu2001b} for more details). Taking the variational derivative of $\Upsilon_\J$ with respect to $|\dot\gamma^{d}_\J)$ and setting it equal to zero results in
\begin{equation}\label{eq:Lagrangian} 
	\frac{\delta\Upsilon_\J}{|\delta\dot\gamma^{d}_\J)}= \Big|2(M)^\J_\rho\gamma_\J\Big) -\Boltz\tau_\J\, \hat G_\J\,\Big|\dot\gamma^d_\J\Big)=0\,,
\end{equation}
where the identity $(X|\,\hat G_\J=\hat G_\J\,|X) $, which follows from the symmetry of the metric $\hat G_\J$ is used and the ``locally perceived  nonequilibrium Massieu operator'' is defined as
\begin{equation}\label{eq:Massieu_operator} 
	(M)^\J_\rho=(S)^\J_\rho-\sum_{q=1}^Q\vartheta_q^\J(C_q)^\J_\rho\,.
\end{equation}
Eq.~(\ref{eq:Lagrangian}) then yields
\begin{equation}\label{eq:SEA_general_supermetric} 
	\Big|\dot\gamma^d_\J\Big)=\frac{1}{\Boltz\tau_\J}\hat G_\J^{-1} 	\Big|2(M)^\J_\rho\gamma_\J\Big)\,,
\end{equation}
Here the Lagrange multipliers $\vartheta_q^\J$ (implicit in $(M)^\J_\rho$) are the solution of the system of equations obtained by substituting Eq.~(\ref{eq:SEA_general_supermetric}) into the conservation constraints,  $\dot{c_q}|_\J=0$. Thus,
\begin{equation}\label{eq:multiplierssuper} 
	\Big((C_\ell)^\J_\rho\gamma_\J\Big|\hat G_\J^{-1}\Big|(M)^\J_\rho\gamma_\J\Big)=0\quad \forall \ell\,.
\end{equation}
Using Cramer's rule, this system can be solved explicitly for the $\vartheta_q^\J$'s to obtain convenient expressions for the $\dot\gamma^{d}_\J$'s as ratios of determinants (as in the original formulations). The $\vartheta_q^\J$'s are nonlinear functionals of $\rho$ that may be interpreted as ``local nonequilibrium entropic potentials'' conjugated with the conserved charges $C_q$. The first two charges are always the identity operator, $C_1\,{=}\,I$, which implements the $\Tr\rho\,{=}\,1$ constraint and the Hamiltonian operator, $C_2\,{=}\,H$, which implements the $\Tr\rho H\,{=}\,\text{constant}$ constraint. The Lagrange multiplier $\vartheta_2^\J$ plays the role of ``local nonequilibrium inverse temperature'' conjugated with the locally perceived energy, and for the stable equilibrium states of the SEA dynamics, it coincides with the  thermodynamic inverse temperature $\Boltz\beta$. 

Additional charges may be considered as generators of the dissipative dynamics. For example,  the vector of particle-number operators $\{C_3\,{=}\,N_1,\dots,C_{3+r-1}\,{=}\,N_r\}$, all commuting with $H$, can be used to implement the number-of-particles conservation constraints  $\Tr\rho N_j\,{=}\,\text{const}$.  The corresponding Lagrange multipliers $\vartheta_i^\J$  play the role of ``local nonequilibrium chemical potentials'' conjugated with the locally perceived number of particles, and for the stable equilibrium states of the SEA dynamics, they converge to  $\Boltz\beta\mu_j$ where the $\mu_j$'s are thermodynamic chemical potentials. In view of the recent interest and results about thermalization with non-commuting ``charges'' \cite{YungerHalpern2016,Murthy2023,YungerHalpern2023,Majidy2023}, it is noteworthy that the requirement that the non-Hamiltonian generators of the motion all commute with the operator $H$ is not necessary for their mean values to be time-invariants of the  SEAQT dissipative term in Eq.~\eqref{eq:SEA_Composite_simp}. It is only necessary for them to be time-invariants of the Hamiltonian nondissipative term. Therefore, the SEAQT equations of motion are already applicable, with no modifications needed, able to model nonequilibrium relaxation towards the thermal state of a quantum system with noncommuting charges.  In the SEAQT framework the operators $C_\ell$ are called the non-Hamiltonian generators of the motion. Other examples include the momentum component operators for a free particle or the magnon operator of a magnetic material.

Following~\cite{Beretta2014}, the ``local  nonequilibrium affinity'' operators are defined as
\begin{equation}\label{eq:noneq_affinity} 
	|\Lambda_\J)= \hat G_\J^{-1/2}\Big|2(M)^\J_\rho\gamma_\J\Big)\,,
\end{equation}
so that the overall rate of entropy production becomes
\begin{equation}\label{eq:entropy_production_SEA_super} 
	\dv{\langle S\rangle}{t}= \sum_{\J=1}^\M\frac{(\Lambda_\J|\Lambda_\J)}{\Boltz\tau_\J}\,,
\end{equation}
where $(\Lambda_\J|\Lambda_\J)$ is the norm of operator $2(M)^\J_\rho\gamma_\J$ with respect to the metric $\hat G_\J^{-1}$ and may be interpreted as the ``degree of disequilibrium'' of subsystem $\J$. Hence, the necessary and sufficient condition for the overall state to be locally nondissipative (no contribution to the overall entropy production from subsystem $\J$) is that operator $2(M)^\J_\rho\gamma_\J$ vanishes.

The metric superoperator $\hat G_\J$ plays a role analogous to the symmetric thermal conductivity tensor $\hat k$ in heat transfer theory. In that context, \(\hat k\) defines the general near-equilibrium linear relationship, $|q'')=-\hat k\,|\nabla T)$, between the heat flux vector $q''$ and the  conjugated ``degree of disequilibrium'' vector, i.e., the temperature gradient $\nabla T$. Here, the SEAQT dissipative term in the equation of motion expresses a more general linear relationship between the local evolution operator $\dot\gamma^d_\J$ and the nonequilibrium Massieu operator, $(S)^\J_\rho - \sum_q\vartheta_q^\J(C_q)^\J_\rho$. This relation is more general, as it holds not only near equilibrium but also anywhere far from equilibrium. In the present quantum modeling context,  it represents the nonlinear SEA extension into the far-nonequilibrium domain of Onsager’s linear near-equilibrium theory, with reciprocity naturally embedded through the symmetry of any metric. As in heat transfer theory where the conductivity tensor for an isotropic material is $\hat k= k\,\hat I$, in the SEAQT formalism an analogous simplification is obtained when $\hat G_\J= \hat I_\J$, the identity operator on $\mathcal{L}(\Hil_\J)$. This corresponds to assuming a uniform Fisher--Rao metric, as is done in early versions of the SEAQT formalism.

In order for the SEAQT equation of motion to be independent of the unitary operators $U$ used (in  $\gamma_\J=\sqrt{\rho_\J}\,U$) to define the generalized square roots of $\rho_\J$,
the choice of the metric superoperator, $\hat G_\J$, is further restricted by the following general assumption: 

\begin{enumerate}[label=\textbf{(SEAQT\arabic*):}, labelsep=4pt,start=4, leftmargin=*]\item
	$\hat G_\J= L_\J^{-1} \hat I_\J$ is assumed, with $L_\J$ a strictly positive, hermitian operator on $\Hil_\J$ and $\hat I_\J$ the identity operator on  $\mathcal{L}(\Hil_\J)$.
\end{enumerate}
From this it follows that: $\hat G_\J|X)~=~|L_\J^{-1} X)$; $\hat G^{-1}_\J|X)~=~|L_\J X)$; $	(X\gamma_\J|\hat G^{-1}_\J|Y\gamma_\J)=\half\Tr\left[ \rho_\J(X^\dagger L_\J Y+Y^\dagger L_\J X)\right]$; and the dissipative terms in Eq.~(\ref{eq:SEA_Composite_simp}) become 
\begin{equation}\label{eq:SEA_general} 
	-\acomm*{\DrhoJ}{\rho_\J}=\frac{2}{\Boltz\tau_\J}\left[L_\J (M)^\J_\rho\rho_\J +\rho_\J (M)^\J_\rho L_\J\right]\,;
\end{equation}
The system of equations that determines the Lagrange multipliers  $\vartheta_q^\J$ in $(M)^\J_\rho$ is then
\begin{equation}\label{eq:multipliers} 
	\Tr\left(\rho_\J\Big[(C_\ell)^\J_\rho L_\J(M)^\J_\rho+(M)^\J_\rho L_\J(C_\ell)^\J_\rho\Big]\right)=0\quad \text{ for } \ell=1,\dots,Q\,;
\end{equation}
where the  dependence on  $\gamma_\J$ is only through the product  $\gamma_\J \gamma_\J^\dagger$, i.e., the local state operator $\rho_\J$; and the (non-negative) entropy production is
\begin{equation}\label{eq:entropy_production_SEA_L} 
	\dv{\langle S\rangle }{t}= 4\sum_{\J=1}^\M\frac{\Tr[\rho_\J (M)^\J_\rho L_\J (M)^\J_\rho]}{\Boltz\tau_\J}\,.
\end{equation}

\begin{enumerate}[label=\textbf{(SEAQT\arabic*):}, labelsep=4pt,start=5, leftmargin=*]\item
	The operator $L_\J$, which determines the local dissipative metric $\hat G_\J= L_\J^{-1} \hat I_\J$, is assumed to commute with  the nonequilibrium Massieu operators $(M)^\J_\rho$, i.e., given the  spectral form $(M)^\J_\rho{\,=\,} \sum_{\iJ=1}^{\dim\Hil_\J}  \miJ\RiJ$ where the $\RiJ$'s are one-dimensional eigenprojectors, the operator $L_\J$ (for every $\J$) can be written as
	\begin{equation}\label{eq:LJ_multipleTimes1} 
		L_\J=\frac{\tau_\J}{4}\sum_{\iJ=1}^{\dim\Hil_\J} \frac{1}{\tauiJ} \RiJ\,,\quad \text{ and } L_\J^{-1}=\frac{4}{\tau_\J}\sum_{\iJ=1}^{\dim\Hil_\J} \tauiJ \RiJ\,,
	\end{equation}
	(the prefactor $\tau_\J/4$  is introduced here to make contact with the notation in previous papers on SEA and SEAQT and to keep operator $L_\J$ dimensionless) 
	so that the SEA dissipator $\DrhoJ$  in the SEAQT Eq.~(\ref{eq:SEA_Composite_simp}) takes the form
	\begin{equation}\label{eq:LJ_multipleTimes2} 
		\DrhoJ=-L_\J(M)^\J_\rho=-(M)^\J_\rho L_\J=-\frac{\tau_\J}{4}\sum_{\iJ=1}^{\dim\Hil_\J} \frac{\miJ}{\tauiJ} \RiJ\,,
	\end{equation}
	and the constraining Eqs. \eqref{eq:multipliers} and the overall entropy production become
	\begin{equation}\label{eq:multipliersL} 
		\Tr\left(\rho_\J\acomm*{\DrhoJ}{(C_\ell)^\J_\rho}\right)=0\quad \text{ for } \ell=1,\dots,Q\,,
	\end{equation} 
	\begin{equation}\label{eq:entropy_production_SEA_R} 
		\dv{\langle S\rangle}{t}= \frac{1}{\Boltz}\sum_{\J=1}^\M\sum_{\iJ=1}^{\dim\Hil_\J} \frac{(\miJ)^2}{\tauiJ} \Tr(\rho \RiJ)\,.
	\end{equation}
	
\end{enumerate}

\section{HE--SEAQT Assumptions for noninteracting Subsystems}\label{sec:heseaqt}
In this appendix, the HE assumptions discussed in Section \ref{sec:CSHE} are merged with the SEAQT assumptions discussed in  Appendix \ref{sec:seaqt} and completed with two additional assumptions to obtain the HE--SEAQT formulation. For simplicity, the minimal set of SEA generators of the motion, i.e., $Q\,{=}\,2$, $C_1\,{=}\,I$ and $C_2\,{=}\,H$, is assumed.
Eq.~\eqref{eq:MassieuJ} with $\Boltz\alpha_\J =\vartheta_1^\J$ and $\Boltz\beta_\J =\vartheta_2^\J$ then becomes
\begin{align}\label{eq:MassieuHESAE} 	 
	(M)^\J_\rho& =  M_\J = S_\J-\vartheta_1^\J I_\J -\vartheta_2^\J H_\J \nonumber\\ &=\Boltz
	\sumKJ [  (\alphaKJ- \vartheta_1^\J/\Boltz)    \PKJ +(\betaKJ- \vartheta_2^\J/\Boltz) \HKJ  ]\nonumber\\ &=\Boltz
	\sumKJ [  (\alphaKJ- \alpha_\J)    \PKJ +(\betaKJ- \beta_\J) \HKJ  ]\, , 
\end{align} 
where,  to simplify the notation in what follows, the superscript HE is dropped and in the last step the Lagrange multipliers, $\vartheta_1^\J=\Boltz\alpha_\J $ and $\vartheta_2^\J=\Boltz\beta_\J $, are renamed and called ``SEA nonequilibrium potentials.''

Since by  Assumption CSHE3 $\rho_\J$ and  $H_\J$ share the same eigenprojectors and degeneracies, the operators $L_\J$,  $M_\J$, and $\rho_\J$ commute with each other, the eigenprojectors $\RiJ$ and $\PiJ$ coincide, and Eqs.~\eqref{eq:SEA_local} and \eqref{eq:SEA_general} can be rewritten as
\begin{equation}\label{eq:SEA_generalHE1} 
	\dv{\rho_\J}{t}~=~-\acomm*{\mathcal{D}^\J_\rho}{\rho_\J}=\frac{4}{\tau_\J\Boltz} M_\J L_\J \rho_\J  =\frac{4}{\tau_\J}\SumKJ [  (\alphaKJ- \alpha_\J)    \PKJ +(\betaKJ- \beta_\J) \HKJ  ]\,  L_\J \rho_\J\,,
\end{equation}
Furthermore, recalling that each $\rho_\J$ is non-singular (Assumption CSHE2),
\begin{equation}\label{eq:SEA_generalHES1} 
	\dv{S_\J}{t}=-\frac{4}{\tau_\J} M_\J L_\J  =-\frac{4\Boltz}{\tau_\J}\SumKJ [  (\alphaKJ- \alpha_\J)    \PKJ +(\betaKJ- \beta_\J) \HKJ  ]\, L_\J \,.
\end{equation}
Eqs. \eqref{eq:multipliers}, which determine the two SEA nonequilibrium  potentials $\alpha_\J$ and $ \beta_\J$, can then be rewritten as
\begin{align}\label{eq:multipliersHE1} 
	\Tr( L_\J \rho_\J M_\J)=	\Tr( L_\J \rho_\J S_\J)-\Boltz\alpha_\J \Tr( L_\J \rho_\J ) -\Boltz\beta_\J \Tr( L_\J \rho_\J H_\J )  &=0\,, \\
	\Tr( L_\J \rho_\J  M_\J H_\J)=	\Tr( L_\J \rho_\J S_\J H_\J)-\Boltz\alpha_\J \Tr( L_\J \rho_\J H_\J  ) -\Boltz\beta_\J \Tr( L_\J \rho_\J H_\J H_\J ) &=0\,,
\end{align}
yielding the solution
\begin{equation}\label{eq:multipliersHE2} 
	\Boltz\alpha_\J = \frac{A_{S}^\J A_{HH}^\J-A_{H}^\J A_{SH}^\J }{A_{I}^\J A_{HH}^\J-A_{H}^\J A_{H}^\J}= \frac{A_{S}^\J}{A_{I}^\J} -	\Boltz\beta_\J\frac{A_{H}^\J}{A_{I}^\J} \,,\qquad
	\Boltz\beta_\J = \frac{A_{I}^\J A_{SH}^\J   - A_{S}^\J A_{H}^\J}{A_{I}^\J A_{HH}^\J -A_{H}^\J A_{H}^\J} \,,
\end{equation}
where the following weighted mean values are defined:
\begin{align}\label{eq:multipliersHEAB1} 
	A_{I}^\J = \Tr( L_\J \rho_\J)\,,\quad
	A_{H}^\J = \Tr( L_\J \rho_\J H_\J)\,,\quad
	A_{S}^\J = \Tr( L_\J \rho_\J S_\J)\,,\\
	A_{HH}^\J =\Tr( L_\J \rho_\J H_\J H_\J)\,,\quad
	A_{SH}^\J = \Tr( L_\J \rho_\J S_\J H_\J)\,.
\end{align}

Finally, the following additional assumptions are assumed:
\begin{enumerate}[label=\textbf{(HE--SEAQT\arabic*):}, labelsep=4pt,start=6, leftmargin=*]
	\item The Hamiltonian operator is time independent; therefore, the energy eigenvalues, eigen-projectors, degeneracies, and sector identity operators $\PKJ$ are as well.
	
	\item Within each HE sector $\smallK,\J$ the relaxation times are the same for all energy eigenlevels, i.e., 
	\begin{equation}
		\tauiKJ=\tauKJ \quad \text{ for all } \iKJ\text{'s and every } \smallK \text{ and } \J \, .
	\end{equation}
\end{enumerate}
It follows that
\begin{equation}
	\frac{4 L_\J}{\tau_\J}=	\SumNJ\frac{1}{\tauiJ} \PiJ = \SumKJ 	\SumiKJ   \frac{1}{\tauiKJ} \PiKJ = \SumKJ \frac{1}{\tauKJ}	\SumiKJ    \PiKJ =\SumKJ \frac{1}{\tauKJ}   \PKJ \,,
\end{equation}
\begin{equation}
	L_\J \rho_\J= \frac{\tau_\J}{4}	\SumKJ \frac{\pKJ}{\tauKJ}   \rhoKJtilde \,,
\end{equation}
and, therefore, recalling  the mutual orthogonality of the sector identities $\PKJ$, Eqs.~\eqref{eq:SEA_generalHE1} and \eqref{eq:SEA_generalHES1} reduce to
\begin{equation}\label{eq:SEA_generalHE} 
	\dv{\rho_\J}{t}~=~-\acomm*{\mathcal{D}^\J_\rho}{\rho_\J}=\SumKJ \frac{\pKJ}{\tauKJ} [  (\alphaKJ- \alpha_\J)    \rhoKJtilde +(\betaKJ- \beta_\J) \HKJ \rhoKJtilde ]\,,
\end{equation}
\begin{equation}\label{eq:SEA_generalHES} 
	\dv{S_\J}{t}=-\Boltz\SumKJ \frac{1}{\tauKJ} [  (\alphaKJ- \alpha_\J)    \PKJ +(\betaKJ- \beta_\J) \HKJ  ] \,,
\end{equation}
\begin{equation}\label{eq:entropy_production_SEAHESJ} 
	\dv{\langle S\rangle}{t}=\Boltz\SumKJ\frac{1}{\tauKJ}\Tr\left(\rhoKJtilde\,\left[  (\alphaKJ- \alpha_\J)    \PKJ +(\betaKJ- \beta_\J) \HKJ  \right]^2\right) \,.
\end{equation}
and the SEA potentials $\alpha_\J$ and $ \beta_\J$ can be written as
\begin{equation}\label{eq:multipliers_generalHE} 
	\Boltz\alpha_\J = \frac{B_{S}^\J B_{HH}^\J-B_{H}^\J B_{SH}^\J }{ B_{HH}^\J-B_{H}^\J B_{H}^\J} =B_{S}^\J- \Boltz\beta_\J B_{H}^\J\,,\qquad
	\Boltz\beta_\J = \frac{ B_{SH}^\J   - B_{S}^\J B_{H}^\J}{ B_{HH}^\J -B_{H}^\J B_{H}^\J} \,,
\end{equation}
where the redefined weighted mean values (Eqs.~\eqref{eq:multipliersHEAB1} with $\tilde\tau_\J=\tau_\J/4A_{I}^\J$ and   $B_{X}^\J= A_{X}^\J/ A_{I}^\J$) are
\begin{align}\label{eq:multipliers_generalHEAB1} 
	\frac{1}{\tilde\tau_\J}= \SumKJ \frac{\pKJ}{\tauKJ}\,,\quad
	B_{H}^\J = \tilde\tau_\J\SumKJ \frac{\langle \HKJ \rangle }{\tauKJ}\,,\quad
	B_{S}^\J = \tilde\tau_\J\SumKJ \frac{\langle \SKJ \rangle }{\tauKJ}\,, \\ 
	B_{HH}^\J =\tilde\tau_\J\SumKJ \frac{\langle \HKJ\HKJ \rangle }{\tauKJ}\,,\quad
	B_{SH}^\J = \tilde\tau_\J\SumKJ \frac{\langle \SKJ\HKJ \rangle }{\tauKJ}\,. \label{eq:multipliers_generalHEAB2}
\end{align}

Consistently with the assumption that the subsystems are noninteracting (interaction Hamiltonian $V=0$), no energy exchange occurs between them. Consequently, each subsystem relaxes independently toward the canonical Gibbs state $\rho_\J^\text{SE}=\text{exp} (-\beta_\J^\text{SE}H_\J)/Z_\J(\beta_\J^\text{SE})$ with inverse temperature $\beta_\J^\text{SE}$ determined uniquely by its initial mean energy $\Tr(\rho_\J(0) H_\J)$.  Since these final temperatures generally differ, the noninteracting subsystems do not reach 
mutual equilibrium. 
Clearly, when no HE decomposition is assumed, $M_\J=1$, Eqs.~\eqref{eq:multipliers_generalHEAB1} and \eqref{eq:multipliers_generalHEAB2} reduce to
\begin{align}\label{eq:multipliers_generalHEAB3} 
	\tilde\tau_\J=  \tau_\J\,,\quad
	B_{H}^\J = \langle H_\J \rangle \,,\quad
	B_{S}^\J = \langle S_\J \rangle \,, \quad
	B_{HH}^\J =\langle H_\J H_\J \rangle\,,\quad
	B_{SH}^\J = \langle S_\J H_\J \rangle\,.
\end{align}


\reftitle{References}
\bibliography{references}

\end{document}